\documentclass[twocolumn]{aastex7}
\pdfoutput=1

\let\tablenum\relax
\usepackage[separate-uncertainty=true, range-phrase=\text{-}, range-units=single,
angle-symbol-over-decimal, retain-explicit-plus, multi-part-units=single]{siunitx}
\usepackage{amsmath}
\usepackage{graphicx}
\usepackage{chemformula}
\usepackage{acro}
\usepackage{layouts}
\usepackage{float}
\usepackage{multirow}
\acsetup{patch/longtable=false}
\usepackage{lipsum}
\usepackage[T1]{fontenc}

\received{}

\shorttitle{}
\shortauthors{}

\DeclareSIUnit{\mag}{mag}
\DeclareSIUnit{\pixel}{pixel}
\DeclareSIUnit{\parsec}{pc}
\DeclareSIUnit{\arcsec}{arcsec}
\DeclareSIUnit{\arcmin}{arcmin}
\DeclareSIUnit{\solarlum}{\mbox{$\mathcal{L}_\odot$}}
\DeclareSIUnit{\solarmass}{\mbox{$\mathcal{M}_\odot$}}
\DeclareSIUnit{\solarmetal}{\mbox{$Z_\odot$}}
\DeclareSIUnit{\year}{yr}
\DeclareSIUnit{\deg}{deg}
\DeclareSIUnit{\erg}{erg}
\DeclareSIUnit{\dex}{dex}
\DeclareSIUnit{\angstrom}{\textup{\AA}}
\DeclareSIUnit{\radius}{\mbox{$R_{25}$}}
\DeclareSIUnit{\spaxel}{spaxel}
\DeclareSIUnit{\jansky}{Jy}

\newcommand{\ha}{H$\alpha$}
\newcommand{\hb}{H$\beta$}
\newcommand{\oiii}{[\ion{O}{3}]}
\newcommand{\nii}{[\ion{N}{2}]}
\newcommand{\oii}{[\ion{O}{2}]}
\newcommand{\sii}{[\ion{S}{2}]}
\newcommand{\siii}{[\ion{S}{3}]}
\newcommand{\hii}{\ion{H}{2}}

\DeclareAcroEnding{possessive}{'s}{'s}

\NewAcroCommand\acg{m}{\acropossessive\UseAcroTemplate{first}{#1}}

\DeclareAcronym{sfr}{short=SFR, long=star formation rate}
\DeclareAcronym{ssfr}{short=sSFR, long=specific star formation rate}
\DeclareAcronym{psf}{short=PSF, long=point spread function}
\DeclareAcronym{ism}{short=ISM, long=interstellar medium}
\DeclareAcronym{dig}{short=DIG, long=diffuse ionized gas}
\DeclareAcronym{ifs}{short=IFS, long=integral field spectroscopy}
\DeclareAcronym{imf}{short=IMF, long=initial mass function}
\DeclareAcronym{sfh}{short=SFH, long=star formation history, long-plural-form=star formation histories}
\DeclareAcronym{sn}{short=S/N, long=signal-to-noise}
\DeclareAcronym{sitelle}{short=SITELLE, long={Spectro-Imageur \`{a} Transform\'{e}e de Fourier pour l'\'{E}tude en Long et en Large des raies d'\'{E}mission}}
\DeclareAcronym{cfht}{short=CFHT, long={Canada-France-Hawaii Telescope}}
\DeclareAcronym{signals}{short=SIGNALS, long={Star-formation, Ionized Gas, and Nebular Abundances Legacy Survey}}
\DeclareAcronym{snr}{short=SNR, long=supernova remnant, long-plural-form=supernovae remnants}
\DeclareAcronym{soba}{short=SOBA, long=scaled OB association}
\DeclareAcronym{wr}{short=WR, long=Wolf-Rayet}
\DeclareAcronym{sne}{short=SN, long=supernova, short-plural-form=SNe, long-plural-form=supernovae}
\DeclareAcronym{ghr}{short=GH\textsc{ii}R, long=giant \hii\ region}
\DeclareAcronym{ifts}{short=iFTS, long=imaging Fourier transform spectroscopy}
\DeclareAcronym{ifu}{short=IFU, long=integral field unit}
\DeclareAcronym{ils}{short=ILS, long=instrument line shape}

\newcolumntype{P}[1]{>{\centering\arraybackslash}p{#1}}

\begin{document}

\title{SIGNALS of Giant \hii\ Regions: A Spatially Resolved Analysis of NGC 604}
\author[0000-0002-9426-7456,gname=Ray,sname=Garner,suffix=III]{Ray Garner, III}
\affiliation{Department of Physics and Astronomy, Texas A\&M University, 578 University Dr., College Station, TX, 77843, USA}
\affiliation{George P.\ and Cynthia W.\ Mitchell Institute for Fundamental Physics \& Astronomy, Texas A\&M University, 578 University Dr., College Station, TX, 77843, USA}
\email{ray.three.garner@gmail.com}

\author[0000-0001-5448-1821,gname=Robert,sname=Kennicutt,suffix=Jr]{Robert C. Kennicutt, Jr.}
\affiliation{Department of Physics and Astronomy, Texas A\&M University, 578 University Dr., College Station, TX, 77843, USA}
\affiliation{George P.\ and Cynthia W.\ Mitchell Institute for Fundamental Physics \& Astronomy, Texas A\&M University, 578 University Dr., College Station, TX, 77843, USA}
\affiliation{Department of Astronomy and Steward Observatory, University of Arizona, 933 N.\ Cherry Avenue, Tucson, AZ 85721, USA}
\email{rck@tamu.edu}

\author[0000-0003-1278-2591,gname=Laurent,sname=Drissen]{Laurent Drissen}
\affiliation{D\'{e}partment de Physique, de G\'{e}nie Physique et d'Optique, Universit\'{e} Laval, Qu\'{e}bec, QC, G1V 0A6, Canada}
\affiliation{Centre de Recherche en Astrophysique du Qu\'{e}bec (CRAQ), Qu\'{e}bec, QC, G1V 0A6, Canada}
\email{laurent.drissen@phy.ulaval.ca}

\author[gname=Carmelle,sname=Robert]{Carmelle Robert}
\affiliation{D\'{e}partment de Physique, de G\'{e}nie Physique et d'Optique, Universit\'{e} Laval, Qu\'{e}bec, QC, G1V 0A6, Canada}
\affiliation{Centre de Recherche en Astrophysique du Qu\'{e}bec (CRAQ), Qu\'{e}bec, QC, G1V 0A6, Canada}
\email{carmelle.robert@phy.ulaval.ca}

\author[0000-0002-5136-6673,gname=Laurie,sname=Rousseau-Nepton]{Laurie Rousseau-Nepton}
\affiliation{Dunlap Institute of Astronomy and Astrophysics, University of Toronto, 50 St.\ George Street, Toronto, ON, M5S 3H4, Canada}
\affiliation{Department of Astronomy \& Astrophysics, University of Toronto, 50 St.\ George Street, Toronto, ON, M5S 3H4, Canada}
%\affiliation{Canada-France-Hawaii Telescope, 65-1238 Mamalahoa Highway, Kamuela, HI 96743, USA}
\email{laurie.rousseau.nepton@utoronto.ca}

\author[0000-0001-5801-6724,gname=Christophe,sname=Morisset]{Christophe Morisset}
\affiliation{Instituto de Astronom\'{\i}a, Universidad Nacional Aut\'{o}noma de M\'{e}xico, Unidad Acad\'{e}mica en Ensenada, Km 103 Carr.\ Tijuana-Ensenada, Ensenada, B.C., C.P.\ 2280, Mexico}
\affiliation{Instituto de Ciencias F\'{\i}sicas, Universidad Nacional Aut\'{o}noma de M\'{e}xico, Av.\ Universidad s/n, 62210 Cuernavaca, Mor., Mexico}
\email{chris.morisset@gmail.com}

\author[0000-0001-5657-4837,gname=Philippe,sname=Amram]{Philippe Amram}
\affiliation{Aix Marseille University, CNRS, CNES, Laboratoire d'astrophysique de Marseille, 38 Rue Fr\'{e}d\'{e}ric Joliot Curie, Marseille 13013, France}
\email{philippe.amram@lam.fr}

\author[gname=Pierre,sname=Martin]{R. Pierre Martin}
\affiliation{Department of Physics and Astronomy, University of Hawaii at Hilo, Hilo, HI, 96720, USA}
\email{rpm33@hawaii.edu}

\author[0009-0006-5612-7336,gname=Emma,sname=Jarvis]{Emma Jarvis}
\affiliation{Dunlap Institute of Astronomy and Astrophysics, University of Toronto, 50 St.\ George Street, Toronto, ON, M5S 3H4, Canada}
\affiliation{Department of Astronomy \& Astrophysics, University of Toronto, 50 St.\ George Street, Toronto, ON, M5S 3H4, Canada}
\email{emma.jarvis@mail.utoronto.ca}

\correspondingauthor{Ray Garner, III}
\email{ray.three.garner@gmail.com}

\begin{abstract}
Observing \aclp{ghr} at fine spatial scales uncovers detailed structures and reveals variations in ionization, abundance, and dynamical properties of ionized gas and the effect of stellar feedback. Using emission line data of M33 observed with \acs{sitelle} as part of the \acl{signals} (\acs{signals}), we present maps of the principal optical emission line ratios for NGC~604, the most luminous \hii\ region in M33. The excitation maps align well with the \ha\ morphology and are clearly related to the location of the central stellar cluster and secondary stellar groups. The maps of ionization-sensitive line ratios show substantial variations across the face of NGC~604. We demonstrate that these variations are unlikely to be due to chemical inhomogeneities but are primarily caused by changes in ionization, which in turn affect the observed line ratios. We present the \ha\ kinematics of the region and connect it to the excitation structure, showing how the dynamic motions influence the spatial distribution of ionized gas. We note two distinct sources identified in these excitation maps: a known supernova remnant and a previously unknown planetary nebula. Such parsec-scale features contribute only a small percentage to the overall light and would remain undetected without the use of high-resolution spatial data. Throughout the paper, we make comparisons to and raise concerns about single-aperture and long-slit spectroscopic measurements of \aclp{ghr}, highlighting the limitations and potential inaccuracies of such methods.

%Although we often simplify them as such, real \hii\ regions are not spherical ionized gas clouds with homogeneous physical properties. To study variations in chemical and ionization properties, high spatial resolution is required, which nearby \aclp{ghr} satisfy well. Using emission line data of M33 observed with \acs{sitelle} as part of the \acl{signals} (\acs{signals}), we present maps of the principle optical emission line ratios for NGC~604, the most luminous \hii\ region in M33. While much of the region falls within the low-density limit, the electron density map reveals clumps of higher-density material, suggesting that geometry may be a contributing factor. The excitation maps align well with the \ha\ morphology and are clearly related to the location of the central stellar cluster and secondary stellar groups. We note two distinct sources identified in these excitation maps: a known supernova remnant and an unknown point source. Maps of abundance- and ionization-sensitive line ratios show substantial variations across the face of NGC~604. We demonstrate that these variations are unlikely to be due to chemical inhomogeneities but instead are primarily caused by changes in ionization. Finally, we present the \ha\ kinematics of the region and connect it to the excitation structure. Throughout the paper, we make comparisons to and raise concerns about single-aperture and long-slit spectroscopic measurements of \aclp{ghr}. 
\end{abstract}

\section{Introduction}

Extragalactic \acfp{ghr} in the nearby universe provide an excellent laboratory for studying star formation processes, the evolution of massive stars, and the properties of the surrounding \ac{ism}. To deduce these properties, emission line spectra are typically used to extract information from individual regions and to study the variation of these properties across multiple \hii\ regions within a galaxy. Such physical properties include the gas conditions (electron temperature and density) from which chemical abundances and ionization conditions are estimated, as well as stellar properties (masses, ages, effective temperatures). These results constitute the main body of knowledge regarding the evolution of disk galaxies \added{\citep[e.g.,][]{aller1942,searle1971,smith1975,pagel1979,diaz1987,vilchez1988,dinerstein1990,kennicutt1998,vanzee1998,rosolowsky2008,berg2015}}. 

In many cases, these emission line spectra are obtained from single-aperture or long-slit measurements of the most intense knots of any given region. Physical properties are then estimated under the assumption that the obtained emission line spectrum is representative of the entire \hii\ region and that any variations are minimal or non-existent. Unfortunately, realistic \hii\ regions are far from the idealized spherical ionized gas clouds with homogeneous physical properties as clearly seen in detailed studies of the nearest Galactic and extragalactic \hii\ regions, e.g., the Orion Nebula \citep{kennicutt2000,sanchez2007,kreckel2024}, 30 Doradus \citep{mathis1985,kennicutt2000,fahrion2024}, and NGC~604 \citep{gonzalezdelgado2000,maizapellaniz2004}, among others. 

The limitation of this assumption has long been recognized \citep[e.g.,][]{diaz1987,rubin1989,castaneda1992}, but technological limits or constraints on observation time have historically precluded progress. A simple solution that has been employed in the past is to sweep an \hii\ region using multiple slits located at different positions, attempting to cover as much surface area as possible \citep[e.g.,][]{kosugi1995,maizapellaniz2004}. However, interpolations are necessary to cover the gaps between slits. Another approach has been the use of narrowband photometry centered on particular emission lines, which, when combined with narrowband photometry of the continuum, results in pure emission line images \citep[e.g.,][]{watkins2017,garner2022}. However, these filters are often $\sim$\SI{100}{\angstrom} wide and blend emission lines together, such as those around \ha. An obvious modern solution is to use \acl{ifs} (\acs{ifs}; \citealt{allingtonsmith2006}) where the use of \acp{ifu} allows for the simultaneous extraction of spatial and spectral information \citep[e.g.,][]{relano2010,sanchez2012}. The drawback is the small FOV, which unfortunately makes very near targets time-expensive as they would require multiple pointings to cover them effectively (although see the SDSS Local Volume Mapper [\citealt{drory2024}] for some promising results). 

Of particular interest to this work is \acl{ifts} (\acs{ifts}; see \citealt{maillard2013} and \citealt{drissen2014} for reviews). This observational tool has several benefits compared to \ac{ifs}. For instance, \ac{ifts} has a much larger FOV than an \ac{ifu} at a fraction of the cost and complexity but at the expense of a shorter spectral coverage and lower sensitivity for continuum and absorption-line sources. As such, these are valuable tools for studying emission line sources such as \hii\ regions \citep{rousseaunepton2018,garner2025,bresolin2025,fernandezarenas2025}, \aclp{snr} (\acsp{snr}; \citealt{duartepuertas2024}), and planetary nebulae \citep{vicensmouret2023}. 

We present here \ac{ifts} observations of NGC~604 collected with the \acl{sitelle} (\acs{sitelle}; \citealt{drissen2019}) as part of the \acl{signals} (\acs{signals}; \citealt{rousseaunepton2019}). NGC~604 is the most luminous \ac{ghr} in M33 (NGC~598) and the second most luminous in the entire Local Group. NGC~604 presents an angular size of $\sim$\ang{;1.6;} and given the large FOV of \ac{sitelle} ($\ang{;11;} \times \ang{;11;}$), we cover the entire region in a single observation. Additionally, at the distance of M33 (\SI{840}{\kilo\parsec}; \added{\citealt{freedman1991,savino2022}}) and an average seeing of $\sim$\ang{;;0.8} results in a high physical resolution of $\sim$\SI{3}{\parsec} allowing us to investigate fine spatial variations in its physical properties. 

NGC~604 has an \ha\ core-halo morphology that shows the impact of the stellar winds of massive stars located in its interior. Unlike in 30~Doradus, where the stars are centrally concentrated, these stars are spread over a large projected area ($\sim$\SI{e4}{\square\parsec} for cluster A in \citealt{hunter1996}) in a structure called a \acl{soba} (\acs{soba}; \citealt{maizapellaniz2004}). The ionizing stellar cluster is composed of a population of at least 200 O-type and \ac{wr} stars \citep{drissen1993,drissen2008,hunter1996} with an age of $\sim$\SI{3}{\mega\year} \citep{hunter1996,gonzalezdelgado2000} on top of an underlying older stellar population of age $\sim$\SI{12}{\mega\year} \citep{eldridge2011}. Photoionized shells and filaments surround this stellar cluster (Figure~\ref{pretty_images}). Previous studies have estimated that the energy of the stellar wind is sufficient to form the central region \citep[e.g.,][]{tenoriotagle2000,maizapellaniz2004,tullmann2008}, while contributions from \acp{snr} expand it into the halo structure \citep[e.g.,][]{yang1996}. 

Most studies of NGC~604 employ long-slit spectroscopy or Fabry-P\'{e}rot observations to analyze the complex kinematics of the gas \citep[e.g.,][]{sabalisck1995,munoztunon1996,yang1996,tenoriotagle2000}. However, very few studies have analyzed the chemical composition and ionization structure of the region. \cite{vilchez1988} obtained an electron temperature of $T_e \simeq \SI{8000}{\kelvin}$, an electron density consistent with the low-density limit\added{\footnote{This is where the \sii\ line ratio asymptotes to a fixed value, below which the line ratio is no longer sensitive to electron density. This corresponds to an \sii\ ratio of $\simeq$1.45 and an electron density of $n_e \lesssim \SI{100}{\per\cubic\centi\metre}$ \citep{osterbrock2006}.}}, and an oxygen abundance of $12 + \log(\mathrm{O/H}) = \num{8.51 \pm 0.03}$, confirmed by \cite{rogers2022}. \cite{maizapellaniz2004} analyzed the excitation properties using multiple long-slit measurements across the face of NGC~604 and found a central excitation structure powered by the main stellar cluster surrounded by secondary excitation pockets produced by smaller stellar groups. Although they were able to produce maps of the common excitation-sensitive emission line ratios, they lacked coverage of the blue \oii$\lambda$3727 lines, which would have allowed for a more comprehensive analysis of the ionization and chemical composition.

In this paper, we analyze \ac{ifts} observations covering the whole face of NGC~604. We organize this paper as follows. In Section~\ref{sec:obs}, we briefly describe the observations and data reduction. In Section~\ref{sec:integrated}, we coadd our data to produce an integrated spectrum of NGC~604. From this spectrum, we make estimates of the \acl{sfr}, mass,  total ionization, and oxygen abundance using strong-line methods. Section~\ref{sec:spatial} presents our spatially-resolved maps, in which we estimate the dust extinction, density, excitation, ionization, and oxygen abundance properties of NGC~604 using commonly-employed strong-line ratios. In Section~\ref{sec:kinematics}, we briefly present maps of the velocity and velocity dispersion and connect it to excitation mechanisms in the previous section. Finally, Section~\ref{sec:important} summarizes the importance of spatially-resolved data in the context of NGC~604, and we present our conclusions in Section~\ref{sec:conclusion}. 

Throughout the text, we use several emission lines and their ratios to unravel the physical parameters of NGC~604. We use the following abbreviations for some of the frequently mentioned emission lines. We denote \nii$\lambda$6584, \sii$\lambda\lambda$6717,6731, \oiii$\lambda$5007, and \oii$\lambda$3727 as \nii, \sii, \oiii, and \oii, respectively, unless otherwise noted. 

\section{Observations \& Data Reduction}\label{sec:obs}

We obtained the data cubes analyzed in this paper in September and October 2017 with the \ac{ifts} \ac{sitelle} \citep{drissen2019} at the Canada-France-Hawaii Telescope. \added{\ac{sitelle} provides the spatially resolved spectra of sources in an $\ang{;11;} \times \ang{;11;}$ field of view with a sampling of \ang{;;0.32}\si{\per\pixel} in three bandpasses: SN3 (\qtyrange{647}{685}{\nano\metre}, with a spectral resolution $R = 2900$), SN2 (\qtyrange{482}{513}{\nano\metre}, $R = 1020$), and SN1 (\qtyrange{363}{386}{\nano\metre}, $R = 1020$). NGC~604 is located in Field 1 of the dataset used to study Wolf-Rayet nebulae \citep{tuquet2024} and supernova remnants \citep{duartepuertas2024} in M33. We refer the interested reader to these papers for more information on the data, but summarize the data collection and reduction process here.}

\added{The raw data, which consists of two complementary 2D interferograms, were transformed into a single fully calibrated spectral data cube with \textsc{orbs}, \acg{sitelle} dedicated data reduction software \citep{martin2015}.} Photometric calibration is secured from images and data cubes of the spectrophotometric standard stars G93-48 and GD71, supplemented by a comparison with the low-resolution spectra of two dozen stars observed with \emph{Gaia} \citep{gaia2016,gaia2023} present in the same field. Wavelength calibration is performed using a high spectral resolution laser data cube, which is improved by measuring the centroid positions of the night sky \ch{OH} emission lines in the SN3 filter, following the procedure described in \citet{martin2018}. The barycentric correction was also applied to the observed spectra. We estimate the uncertainty on the relative velocity in the SN3 data cube across the entire field of view to be less than \SI{2}{\kilo\metre\per\second} \citep[see also][]{martin2021}. 

\added{Spectral data from \ac{sitelle} shows (Figure~\ref{int-spec}) that each line is surrounding by oscillating sidelobes, alternating between positive and negative values. This is because, as described by \cite{martin2016}, \acg{sitelle} \ac{ils} is a sinc function. However, any line broadening caused by turbulent motion, an expanding bubble, or a velocity gradient along an extended aperture will transform the natural \ac{ils} into a sincgauss function, the convolution of a sinc and a Gaussian function. \textsc{orcs}, \acg{sitelle} analysis software suite \citep{martin2016}, is thus used to fit the observed line profiles for each pixel to produce maps or single values of line intensities, radial velocity, and velocity dispersion as well as their uncertainties. 

\textsc{orcs} simultaneously fits all the lines within a given data cube, but allows the user to either force identical velocities and velocity dispersions for all the lines, to let them be completely independent of each other, or to select groups of lines with the same values. The maps and integrated spectra presented here were obtained with the first option. To produce the 2D maps, we chose to subtract a common background to all pixels in a given cube, selected from a region without any obvious nebular contribution.  }

\section{Integrated Properties}\label{sec:integrated}

\begin{figure*}
\includegraphics[width=\textwidth,keepaspectratio]{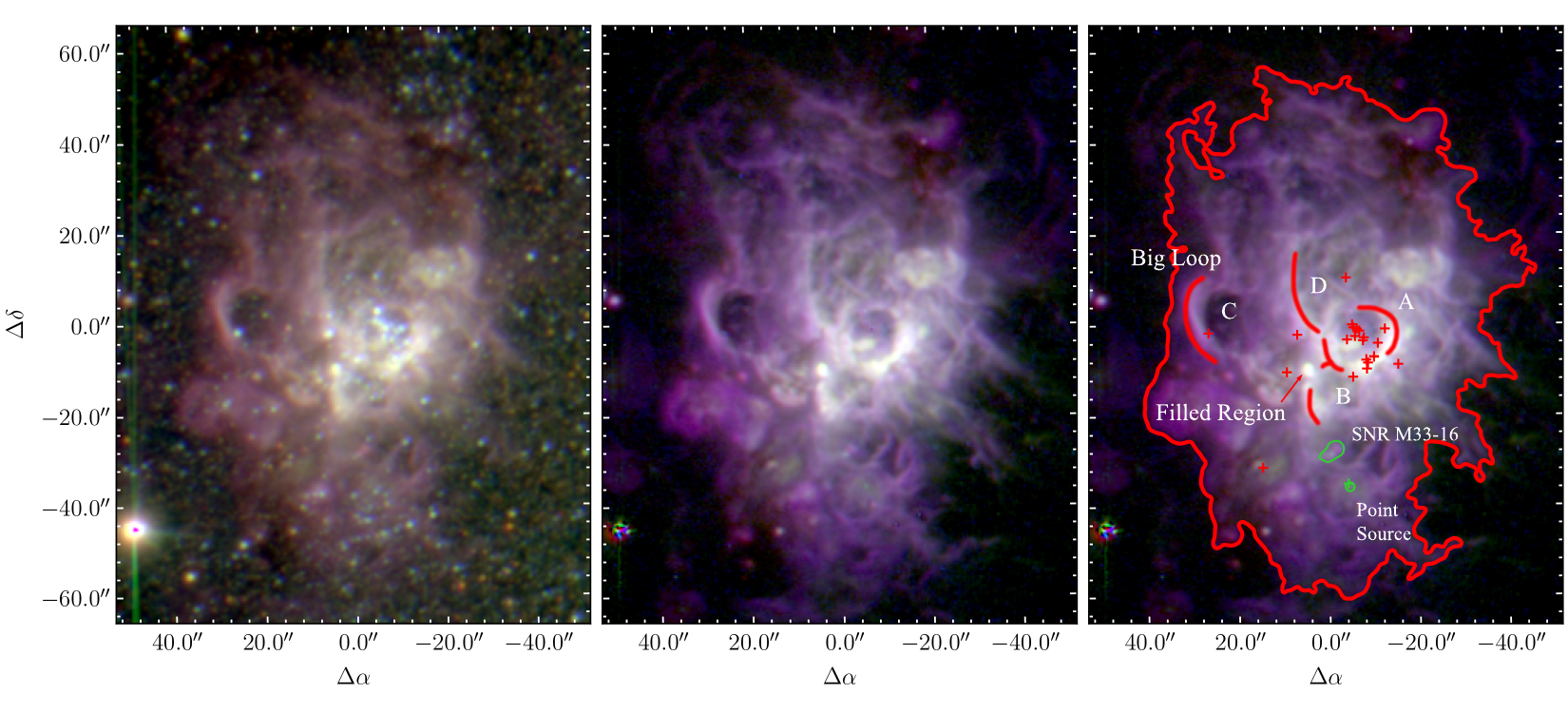}
\caption{Left: \acg{sitelle} deep image of NGC~604. For each pixel, the information from all three filters were summed: SN1 (blue), SN2 (green), and SN3 (red). Middle: an RGB image showing only the gas emission lines \oii$\lambda$3727 (blue), \oiii$\lambda$5007 (green), and \ha\ (red). Right: the same image as in the middle panel with particular features annotated. The outer red contour defines NGC~604 as is ued for the integrated spectrum and corresponds to a flux level of \SI{2e-17}{\erg\per\second\per\square\centi\metre}. The green contours define a known supernova remnant and a point source (see the text for details). Bubbles noted by \cite{maizapellaniz2004} are marked with letters. The red crosses are the sites of \acl{wr} stars and OB stars from \cite{drissen2008} and \cite{bruhweiler2003}, respectively. The green cross near the point source is a star detected in near-infrared imaging by \cite{farina2012}. The coordinate system is the offset from the center coordinate \ang[angle-symbol-degree=\textsuperscript{h},
        angle-symbol-minute=\textsuperscript{m},
        angle-symbol-second=\textsuperscript{s}]{1;34;33.11}, \ang{+30;47;6.8}. North is up and east is to the left.}
\label{pretty_images}
\end{figure*}

Before discussing the emission line maps of NGC~604, we first analyze the spectrum for the entire \hii\ region. To extract the spectrum, we define a custom region around NGC~604 above an \ha\ flux level of \mbox{\SI{2e-17}{\erg\per\second\per\square\centi\metre}}, i.e., the outermost red contour in Figure~\ref{pretty_images}. We chose this flux level to encompass both the bright core of NGC~604 and most of the diffuse halo. We apply this region to each of our three data cubes, extracting and coadding the signal from each of the $\sim$\num{61870} pixels across an area of \SI{1.825}{\square\arcmin} to produce the integrated spectrum.

\begin{figure*}
\includegraphics[width=\textwidth,keepaspectratio]{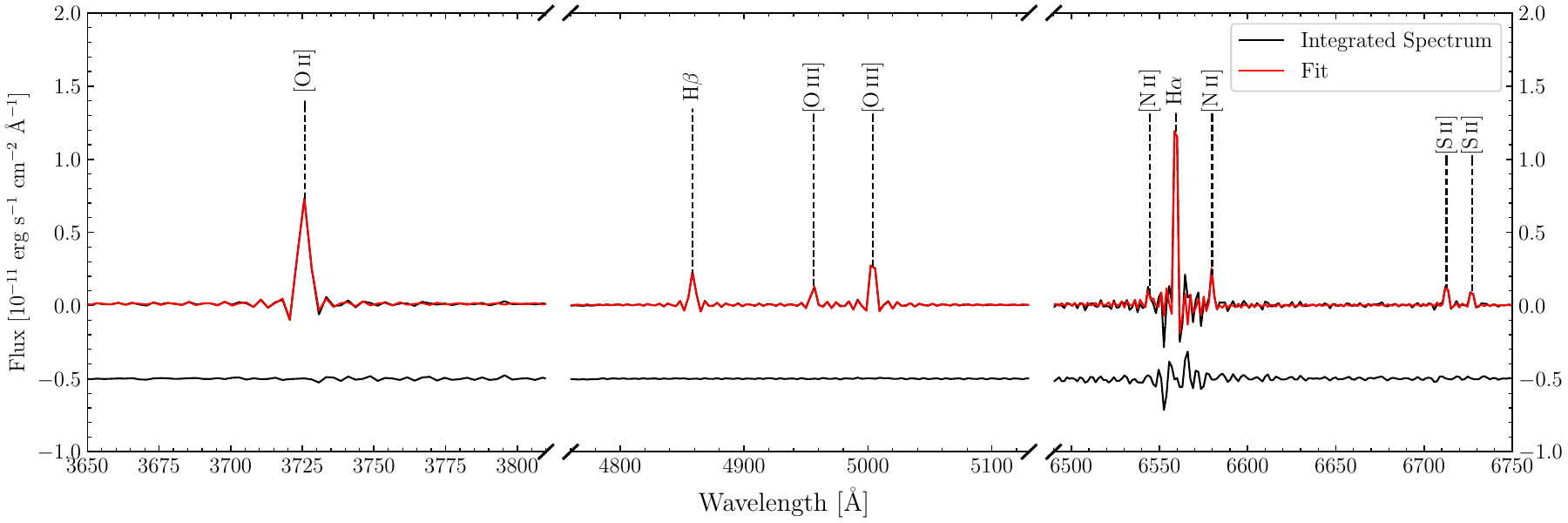}
\caption{The integrated spectrum of NGC~604 with the apparent strong lines indicated. The black line represents the spectrum, while the red line is the fit performed by \textsc{orcs} using a sincgauss function. The black line below the fit is the fit residuals. Please note the changing wavelength along the broken $x$-axis.}
\label{int-spec}
\end{figure*}

\added{Figure~\ref{int-spec} shows the integrated spectrum with its strong optical emission lines in black. We fit the emission lines using \textsc{orcs} and a sincgauss function, tying the velocities of emission lines in each data cube. The resulting fit is shown in red with the residuals below in black. While the SN1 and SN2 data are fit well, there are strong residuals around the \ha\ and \nii$\lambda\lambda$6548,6584 lines in the SN3 data cube. This is likely due to the way the data was collected: the SN1 and SN2 data cubes were obtained in a single night, but the SN3 cube was split across three separate nights. We suspect this introduced a small error in the optical path difference in the interferometer, mimicking a phase error and thus producing the asymmetric line shape. The impact of a phase error on the \ac{ils} and the resulting flux was studied by \cite{martin2021}--see their Appendix A, Equation~A3, and Figure~A1. The observed asymmetry is akin to a phase shift of $\sim$$\pi/6$, resulting in a flux error of \qtyrange{3}{4}{\percent}. Incorporating the phase shift still leaves strong residuals, and these are likely caused by multiple unresolved velocity components, motivating a follow-up campaign to observe NGC~604 at higher resolution (see also Section~\ref{sec:kinematics}).}

Table~\ref{int-lines} lists the emission line fluxes derived from a fit, using a phase-shifted sincgauss function for the SN3 data cube and a standard sincgauss function for the SN1 and SN2 data cubes, normalized to the observed flux of \hb\ in units of \SI{e-12}{\erg\per\second\per\square\centi\metre}. The associated $1\sigma$ errors are solely due to the statistical uncertainty in the measurement of the line fluxes. \added{We checked the robustness of these fits by checking the observed line ratios of \oiii$\lambda$5007/$\lambda$4959 and \nii$\lambda$6584/$\lambda$6548, which should be $2.89$ and $2.94$, respectively, using the transition probabilities of \cite{froesefischer2004} and the collision strengths of \cite{tayal2011} for \ch{N^+} and \cite{storey2014} for \ch{O^{2+}}. While the \oiii\ line ratio matches the predicted ratio within the uncertainties, the \nii\ ratio did not. This is likely caused by the combination of the phase error and multiple unresolved velocity components. Due to the resolution of the SN3 data cube ($R \sim 2900$), instead of fitting multiple components, we elect to fix the \nii\ line ratio to the theoretical value. The data reported in Table~\ref{int-lines} reflects this decision. }

We corrected the observed line fluxes for reddening using the Balmer decrement according to the \cite{calzetti2000} extinction curve, assuming $R_V=3.1$. We assume Case B recombination \added{($T_e = \SI{e4}{\kelvin}$, $n_e = \SI{e2}{\per\cubic\centi\metre}$;  \citealt{osterbrock2006}) and an intrinsic Balmer decrement of $\text{\ha/\hb} = 2.86$ using the atomic data of \cite{storey1995}} while dereddening the observed fluxes. Table~\ref{int-lines} also shows these reddening-corrected emission line fluxes for the integrated spectrum. Our derived dust extinction of $A_V \simeq 0.3$ is consistent with the values reported by \citet[][\SI{0.24}{\mag}]{maizapellaniz2004} and  \citet[][\SI{0.37}{\mag}]{relano2009}.

\begin{deluxetable}{l c c}
\tablecaption{Integrated Line Fluxes for NGC~604 \label{int-lines}}
\tablehead{\colhead{Line} & \colhead{Observed} & \colhead{Corrected}}
\startdata
\oii$\lambda$3727 & \num{3.040 \pm 0.067} & \num{3.398 \pm 0.468} \\
\hb\ $\lambda$4861 & \num{1.000 \pm 0.018} & \num{1.000 \pm 0.116} \\
\oiii$\lambda$4959 & \num{0.551 \pm 0.014} & \num{0.546 \pm 0.063} \\
\oiii$\lambda$5007 & \num{1.576 \pm 0.026} & \num{1.558 \pm 0.177} \\
\nii$\lambda$6548 & \num{0.171 \pm 0.018} & \num{0.153 \pm 0.022} \\
\ha\ $\lambda$6563 & \num{3.198 \pm 0.090} & \num{2.859 \pm 0.288} \\
\nii$\lambda$6584 & \num{0.502 \pm 0.052} & \num{0.448 \pm 0.063} \\
\sii$\lambda$6717 & \num{0.331 \pm 0.057} & \num{0.294 \pm 0.056} \\
\sii$\lambda$6731 & \num{0.233 \pm 0.057} & \num{0.207 \pm 0.054} \\
 & & \\
$F(\mathrm{H}\beta) \, \lambda 4861$ & \num{9.900 \pm 0.130} & \num{13.648 \pm 1.121} \\
$E(B-V)$ & \num{0.096 \pm 0.024} & \\
$A_V$ & \num{0.295 \pm 0.075} & \\
\enddata
\tablecomments{The first column corresponds to the emission line identification. The second column reports the observed line fluxes normalized to the observed, background-corrected \hb\ flux, while the third column reports the reddening-corrected line fluxes normalized to the reddening-corrected \hb\ flux. The \hb\ fluxes are in units of \SI{e-12}{\erg\per\second\per\square\centi\metre}. In the case of the \nii\ lines, these have been fixed to their theoretical ratio of $2.94$ (see text).}
\end{deluxetable}

\subsection{Star Formation Rate \& Mass}

Our measurement of the \ha\ flux allows us to estimate the \acf{sfr} of the entire region and compare it against other \acp{sfr} derived in the literature. In order to determine the \ac{sfr} from the extinction-corrected \ha\ flux, we first convert it to an equivalent \ha\ luminosity assuming a distance of \SI{840}{\kilo\parsec} \citep{freedman1991} and apply the calibration of \cite{murphy2011}: 
\begin{equation}
	\text{SFR} [\si{\solarmass\per\year}] = \num{5.37e-42}L_{\text{\ha}} .
\end{equation}
This calibration assumes solar metallicity and a \cite{kroupa2001} \ac{imf} for a continuous mode of star formation at equilibrium. 

Our observed background-corrected \ha\ luminosity of NGC~604 is \SI[separate-uncertainty-units=bracket]{3.295 \pm 0.192 e39}{\erg\per\second}, in good agreement with measurements using \emph{HST} \ha\ imaging \citep{kennicutt1984,churchwell1999,gonzalezdelgado2000,bosch2002,relano2009}. This luminosity is comparable to the total \ha\ luminosity of 30~Doradus ($L_{\text{\ha}} = \SI{5.13e39}{\erg\per\second}$; \citealt{kennicutt1986}). The observed \ha\ luminosity corresponds to a \ac{sfr} of \SI{0.018 \pm 0.001}{\solarmass\per\year}. 

We also calculate the total number of ionizing photons required to produce the observed \ha\ luminosity and compare that to models of the number of Lyman continuum photons produced by a given stellar type. The minimum number of ionizing photons required to power an \hii\ region is related to the \ha\ luminosity by
\begin{equation}
	Q_0 [\si{\per\second}] = (\SI{7.31e11}{\per\erg})L_{\text{\ha}}
\end{equation}
under Case B recombination \citep{osterbrock2006}. This results in an ionizing photon rate of $\log Q_0 = \num{51.38 \pm 0.03}$. This is equivalent to approximately 130 O5V stars, 90 O5III stars, or 60 O5I stars \citep{martins2005}. This is also similar to previous measurements (51.5; \citealt{kennicutt1984}).

%%% Compare with Kennicutt 1984 ApJ

Observationally, the ionizing stellar cluster is composed of at least 200 O-type stars \citep{hunter1996} and some number of \ac{wr} stars \citep{dodorico1981,rosa1982,diaz1987,drissen1993,drissen2008}. Of these stars, there are only 40 OB stars and approximately 10 \ac{wr} stars with accurate classifications \citep{bruhweiler2003,drissen2008}. Using these classifications and the ionizing fluxes presented by \citet[][O-types]{martins2005} and \citet[][B-types and WR stars]{smith2002}, we calculated a total ionizing flux from the observed stars of $\log Q_{\text{obs}} = 50.72$. Thus, the currently classified stars account for only $\sim$\SI{22}{\percent} of the necessary ionizing flux to produce the \ha\ emission we observe. However, the number of stars with accurate classifications is only \SI{25}{\percent} of the total number of O-type stars \citep{hunter1996}. It is possible that the stars missing classification make up for the missing ionizing flux. 

%%% Volume * n_e * ff 
%%% ff = (n_rms/n_e)^2
%%% n_e = rms/sqrt(ff)

%%% log ne = 2.0 pm 0.5 -> ne = 100 pm 121
%%% nrms = 1.5 pm 0.6
%%% ff = 0.02 pm 0.08  ### 0.0002 pm 0.0006
%%% M = 2.7e5 pm 11.8e5 solar mass

The most common luminosity-based technique used in the literature to estimate ionized gas masses is based on the luminosity of \hb\ and assumes that the emission is dominated by recombination and neglects other emission processes. This is essentially ``photon counting'' and allows for the recombination coefficient to relate the number of photons to the number of hydrogen atoms, and thus the ionized gas mass, i.e., $M_{\text{H\textsc{ii}}} \propto L/n_e$. We utilize the expression derived by \cite{revalski2022},
\begin{equation}\label{eq:mass}
	M_{\text{H\textsc{ii}}} = \left(\frac{L_{\text{H}\beta}}{j_{\text{H}\beta}}\right)\left(\frac{m_{p,\text{eff}}}{n_{p,\text{eff}}}\right).
\end{equation}
Here, $L_{\text{H}\beta}$ is the luminosity of the \hb\ emission line and $j_{\text{H}\beta}$ is the emissivity of \hb. The emissivity has a mild dependence on the electron temperature and electron density of the gas \citep{osterbrock2006}. For the electron temperature, we take the mean value obtained by \cite{esteban2009}, \mbox{$T_e = \SI{8680 \pm 200}{\kelvin}$}. For the electron density, we use an upper limit of \SI{10}{\per\cubic\centi\metre} since the measured \sii$\lambda$6717/\sii$\lambda$6731 of $\simeq$1.45 lies within the low-density limit \citep{osterbrock2006}. We use \texttt{PyNeb} \citep{luridiana2015} and the atomic data of \cite{storey1995} to calculate the emissivity under these conditions, finding that \mbox{$j_{\text{H}\beta} = \SI[separate-uncertainty-units=bracket]{1.394 \pm 0.028 e-25}{\erg\cubic\centi\metre\per\second}$}.

Meanwhile, the other quantities in Equation~\ref{eq:mass} are the effective hydrogen mass, \mbox{$m_{p,\text{eff}} = \mu m_p \approx 1.4m_p$}, and the effective electron density, \mbox{$n_{p,\text{eff}} \approx 1.1n_e$}, both of which take into account elements other than hydrogen. Substituting these quantities into Equation~\ref{eq:mass} gives a total ionized gas mass for NGC~604 of \SI[separate-uncertainty-units=bracket]{8.9 \pm 0.7 e5}{\solarmass}. This is in excellent agreement with the mass estimated by \cite{kennicutt1984} of \SI{7e5}{\solarmass}. In the latter, they assumed a spherically symmetric shell model to estimate the emission measure. While NGC~604 is clearly not a sphere (Figure~\ref{pretty_images}), our agreement shows the robustness of this method to estimate gas masses.  

\added{Of the two observed quantities in Equation~\ref{eq:mass}, the electron density is the least certain. Our mass estimate (as well as that of \citealt{kennicutt1984}) does not take into account any uncertainties on the electron density and should be treated as a lower limit. Other line ratios, such as \oii$\lambda$3729/$\lambda$3726, [\ion{N}{1}]$\lambda$5200/$\lambda$5198, [\ion{Cl}{3}]$\lambda$5517/$\lambda$5537, and [\ion{Ar}{4}]$\lambda$4711/$\lambda$4740, are sensitive to the electron density \citep{kewley2019} and have been measured in NGC~604 \citep{esteban2009}. These density indicators often result in higher densities than the \sii\ ratio, which would naturally decrease the ionized gas mass estimated here. However, [\ion{Cl}{3}] and [\ion{Ar}{4}] trace higher excitation states and thus densities derived from these lines need not be the same as those derived from the low excitation \oii\ and \sii\ lines depending on the density structure of the nebula \citep{rubin1989,wang2004}. Thus, these electron densities are sensitive to where they are measured, especially in the case of long-slit spectra which might preferentially target easily observed high excitation portions of a \ac{ghr} \citep[e.g.,][]{vilchez1988,esteban2009,rogers2022}. }

Finally, we estimate the stellar mass in NGC~604 by integrating the \ac{imf} over the mass range from \qtyrange[range-phrase={ \text{to} }]{0.1}{100}{\solarmass} following the method described by \cite{relano2005}. We summarize the method here but refer the interested reader to the discussion of \cite{relano2005} for more details. 

The total stellar mass is defined as
\begin{equation}\label{eq:stellar_mass}
	M_\ast = \int_{1 \, \mathcal{M}_\odot}^{100 \, \mathcal{M}_\odot} m \Phi(m) \, \mathrm{d}m,
\end{equation}
where $\Phi(m) = Am^{-2.7}$ for a \cite{kroupa2001} \ac{imf}. To determine the proper normalization factor, $A$, we make a first-order estimate of $M_\ast$ by calculating the number of O5V stars needed to produce the measured \ha\ luminosity and multiplying that by the mass of an O5V star, \SI{37.28}{\solarmass} \citep{martins2005}. We substitute this into Equation~\ref{eq:stellar_mass} with the upper and lower limits replaced by the masses of an O4V (\SI{46.16}{\solarmass}) and O6V star (\SI{31.73}{\solarmass}) star, respectively. The normalization factor, $A$, is then solved for, and the \ac{imf} can be integrated over the full mass range to give a more accurate estimate of $M_\ast$. 

For NGC~604, we derive a stellar mass of \mbox{$M_\ast = \SI[separate-uncertainty-units=bracket]{2.3 \pm 0.3 e5}{\solarmass}$}, comparable to that estimated by \citet[][\SI{2e5}{\solarmass}]{yang1996} who assumed a \cite{salpeter1955} \ac{imf}. Here, we propagated the uncertainty from the ionizing flux required to produce the observed \ha\ luminosity. The true uncertainty on this quantity is likely much bigger. A \cite{kroupa2001} \ac{imf} may not be the most reasonable form to describe the stellar content of NGC~604. \cite{drissen1993} determined the slope of the \ac{imf} to be $\alpha = 1.88$ in the range \qtyrange{15}{60}{\solarmass}. This slope is extremely flat compared to a \cite{salpeter1955} \ac{imf}. Assuming that this slope extends to lower masses, the estimated stellar mass decreases by $\sim$\SI{75}{\percent}. Given that the optical imaging survey of \cite{drissen1993} could not detect heavily-extinguished stars \citep{maizapellaniz2004,relano2009,miura2010}, the estimated stellar mass is likely a lower limit at best.

\subsection{Excitation and Abundance}\label{sub:oxy}

Flux ratios between the various emission lines can be used to derive the average properties of the ionized gas in NGC~604. In Table~\ref{int-ratios}, we show the principal diagnostic line ratios derived from the integrated spectrum as well as the physical parameters derived from them. 

\begin{deluxetable*}{l l c}
\tablecaption{Integrated Line Ratios and Physical Properties of NGC~604 \label{int-ratios}}
\tablehead{\colhead{Parameter} & \colhead{Ratio} & \colhead{Value}}
\startdata
Line Ratios & $\log\text{\nii}\lambda6584/\text{\ha}$ & \num{-0.805 \pm 0.056}  \\
&$\log\text{\sii}\lambda\lambda\text{6717,6731}/\text{\ha}$ & \num{-0.757 \pm 0.068}  \\
&$\log\text{\oiii}\lambda5007/\text{\hb}$ & \num{0.193 \pm 0.049}  \\
&$\log\text{\oii}\lambda3727/\text{\hb}$ & \num{0.531 \pm 0.060}  \\
&$\log R_{23}$ & \num{0.741 \pm 0.047}  \\
&$\log \text{O}_{32}$ & \num{-0.339 \pm 0.059}  \\
&$\log\text{\nii}\lambda6584/\text{\oii}\lambda3727$ & \num{-0.880 \pm 0.069}  \\
&$\log\text{\nii}\lambda6584/\text{\sii}\lambda\lambda6717,6731$ & \num{-0.048 \pm 0.081} \\
&\sii$\lambda$6717/\sii$\lambda$6731 & \num{1.421 \pm 0.433} \\
Electron Density & $n_e$ [\si{\per\cubic\centi\metre}] & $<$10 \\
Oxygen Abundance & $R_{23}$ \citep{mcgaugh1991} & \num{8.63 \pm 0.17} \\
& $R_{23}$ \citep{kobulnicky2004} & \num{8.79 \pm 0.16} \\
& $R_{23}$ \citep{pilyugin2005} & \num{8.24 \pm 0.13} \\
& O3N2 \citep{marino2013} & \num{8.32 \pm 0.18} \\
& N2S2H$\alpha$ \citep{dopita2016} & \num{8.51 \pm 0.10} \\
& N2O2 \citep{kewley2019} & \num{8.59 \pm 0.24} \\
Ionization Parameter & O$_{32}$ \citep{kobulnicky2004} & \num{-2.95 \pm 0.15} \\
& O$_{32}$ \citep{kewley2019} & \num{-2.92 \pm 0.07} \\
\enddata
\tablecomments{The oxygen abundance entries indicate what strong-line ratios are used, i.e., $R_{23}$, N2O2, etc., and which calibration is applied given that strong-line ratio. Similarly, the ionization parameter, $\log\mathcal{U}$, entries indicate the strong-line ratio index and the particular calibration used. In both cases, the reported uncertainties take into account the observational uncertainties on the individual line ratios as well as the statistical uncertainties on the calibrations. In all cases, the uncertainties are dominated by the statistical uncertainties.}
\end{deluxetable*}

Several emission line diagnostics are typically used to determine the ionization source for galaxies and emission line regions \citep{baldwin1981,veilleux1987}. The values of \mbox{$\log(\text{\nii}/\text{\ha}) \simeq -0.8$}, \mbox{$\log(\text{\sii}/\text{\ha}) \simeq -0.8$}, and \mbox{$\log(\text{\oiii}/\text{\hb}) \simeq 0.2$} corresponds to the expected values for \hii\ regions photoionized by hot OB stars. These values are consistent with those reported by \cite{maizapellaniz2004}, who used \emph{HST} imaging that covered the entire region. In contrast, our values differ significantly from results based on slit spectroscopy \citep{vilchez1988,esteban2009,rogers2022}, which report higher excitation ratios. This discrepancy likely arises because slit spectra tend to target individual bright knots and do not include the extended diffuse halo. For instance, the spectra in \cite{rogers2022} focus on the brightest, highest-excitation knot in NGC~604, which is likely a compact \hii\ region itself (Figure~\ref{pretty_images}).

The electron density, $n_e$, is estimated from the \sii$\lambda$6717/\sii$\lambda$6731 ratio. Unfortunately, our \sii\ ratio of $\sim$\num{1.42} lies within the low-density limit \citep{osterbrock2006}. This implies densities $<$\SI{10}{\per\cubic\centi\metre}, consistent with all previously estimated densities \citep{vilchez1988,maizapellaniz2004,esteban2009,rogers2022}. Since our filters do not cover the \oiii$\lambda$4363 auroral line, we cannot make an electron temperature measurement. In order to explore the abundances of NGC~604, we will rely on strong-line calibrations but will take the mean value obtained by \cite{esteban2009}, $T_e = \SI{8680 \pm 200}{\kelvin}$, as our standard value when necessary.

Various strong-line emission diagnostics are often used to estimate the oxygen abundance and ionization parameter of an \hii\ region. Among these, the most widely used is the $R_{23}$ index \citep{pagel1979}, defined as
\begin{equation}\label{eq:r23}
	R_{23} = \frac{\text{\oii}\lambda3727 + \text{\oiii}\lambda\lambda\text{4959,5007}}{\text{\hb}},
\end{equation}
which correlates with oxygen abundance but is known to suffer from several limitations (see, e.g., \citealt{mcgaugh1991,pilyugin2001,kewley2002,kewley2008,kewley2019}). 

A well-known issue with the $R_{23}$ indicator is that it is double-valued with respect to oxygen abundance: a given $R_{23}$ value corresponds to two possible solutions, a metal-poor and a metal-rich branch. To resolve this degeneracy, we follow the recommendation of \cite{kewley2008} and use the \nii/\oii\ ratio as a secondary diagnostic, which places NGC~604 on the metal-rich branch. To calculate oxygen abundances for the integrated spectrum using $R_{23}$, we apply three calibrations: the \cite{mcgaugh1991} photoionization model-based calibration parameterized by \cite{kuziodenaray2004}; the \cite{kobulnicky2004} photoionization model-based calibration; and the empirical calibration of \cite{pilyugin2005}. 

Another complication is that $R_{23}$ has a secondary dependence on the ionization parameter, $\mathcal{U}$, which is the ratio of the number density of ionizing photons to the number density of hydrogen atoms \citep{kewley2002}. Although not directly measurable, $\mathcal{U}$ can be estimated through ionization-sensitive ratios \citep[e.g.,][]{kewley2019} such as
\begin{equation}\label{eq:o32}
	\mathrm{O}_{32} = \frac{\text{\oiii}\lambda\lambda\text{4959,5007}}{\text{\oii}\lambda3727}.
\end{equation}
All three $R_{23}$ calibrations used here account for ionization to some degree. However, only the \cite{kobulnicky2004} calibration explicitly returns a value for $\log\mathcal{U}$ after an iterative procedure that converges both O/H and $\log\mathcal{U}$. We also include the $\log\mathcal{U}$ calibration of \cite{kewley2019}, which is also an iterative procedure. We use the ionization-insensitive abundance indicator of N2O2 (see below) in this routine.

While $R_{23}$ has long served as a standard abundance diagnostic, its degeneracy and sensitivity to ionization has motivated the development of alternative indicators that exhibit monotonic behavior with metallicity. \added{These include the empirical O3N2 index \citep{marino2013}, the photoionization model-based N2S2\ha\ index \citep{dopita2016}, and the photoionization model-based N2O2 index \citep{kewley2019}}, defined as
\begin{align}
	\text{O3N2} &= \log\left(\frac{\text{\oiii}\lambda5007}{\text{\hb}}\right) - \log\left(\frac{\text{\nii}\lambda6584}{\text{\ha}}\right), \label{eq:o3n2} \\
	\text{N2S2\ha} &= \log\left(\frac{\text{\nii}\lambda6584}{\text{\sii}\lambda\lambda\text{6717,6731}}\right) \label{eq:n2s2ha} \\
	&\qquad - 0.264\log\left(\frac{\text{\nii}\lambda6584}{\text{\ha}}\right), \nonumber \\
	\text{N2O2} &= \log\left(\frac{\text{\nii}\lambda6584}{\text{\oii}\lambda3727}\right) \label{eq:n2o2}.
\end{align}
Each of these has its limitations, particularly their dependencies on ionization parameter or elemental abundance ratios (e.g., N/O, S/O). For a broader discussion of these diagnostics, see the reviews by \cite{maiolino2019} or \cite{kewley2019}. 

We estimate uncertainties in the derived abundances using a Monte Carlo method. We recalculated each diagnostic 500 times using Gaussian-distributed line fluxes with widths equal to their measurement uncertainties. However, each calibration comes with its own statistical uncertainty. We add these two uncertainties together in quadrature; in all cases, the statistical uncertainty of the calibration dominates. We note that \cite{dopita2016} did not report a statistical uncertainty on their N2S2\ha\ calibration; the reported uncertainty only takes into account the observational uncertainties. Additionally, since the $\log\mathcal{U}$ calibration of \cite{kobulnicky2004} is found iteratively with the oxygen abundance, we assume it has the same statistical uncertainty as the oxygen abundance. The resulting oxygen abundances and ionization parameters and their uncertainties are reported in Table~\ref{int-ratios}. 

Taking a average of these abundance estimates weighted by their uncertainties gives an oxygen abundance of $12 + \log(\mathrm{O/H}) = \num{8.48 \pm 0.07}$. The similarity of this averaged abundance to the N2S2\ha\ abundance might indicate that this abundance is weighing the average. We tested this by removing this abundance estimate from our average; the resulting average was shifted down by only \SI{0.01}{\dex}. Thus, we see that NGC~604 likely has a slightly subsolar oxygen abundance ($\sim$$0.6Z_\odot$) and a mild ionization parameter. The average oxygen abundance estimated here is very similar to that from auroral line electron temperature measurements: \num{8.51 \pm 0.03} \citep{vilchez1988} and \num{8.50 \pm 0.03} \citep{rogers2022}. 

These abundance estimates provide a baseline for interpreting the spatially-resolved structure across NGC~604. As we will show in later sections, the apparent consistency in some indicators masks more complex spatial variations in others, raising key questions about the reliability of different diagnostics when applied to extended nebulae. In particular, this analysis connects directly to the long-standing strong-line abundance discrepancy problem \citep{kewley2008,moustakas2010}, in which different strong-line indicators yield systematically different oxygen abundances, even within the same region \citep[e.g.,][]{relano2010,monrealibero2011,mao2018}.

\section{Spatially-Resolved Properties}\label{sec:spatial}

Before presenting the emission-line ratio maps of NGC~604, we first present a brief overview of the nebular structure of NGC~604. Much of this structure has been discussed extensively in the past \citep{tenoriotagle2000,maizapellaniz2004}; we will not repeat what has already been said. Instead, we highlight several morphological features that will guide the remainder of the discussion. 

The right panel of Figure~\ref{pretty_images} points out several morphological features of NGC~604. In general, the structure of NGC~604 consists of a central, bright, high-excitation region surrounded by a dimmer, low-excitation halo. The central region is composed of a shell surrounding a central cavity (labeled A following the nomenclature of \citealt{maizapellaniz2004}) centered on coordinates (\ang{;;-7},\ang{;;-3}). Many of the most massive ionizing stars reside in cavity A \citep{drissen1993,hunter1996,bruhweiler2003,drissen2008,farina2012}. Two other cavities are visible nearby: cavity B to the south and cavity D to the north. A small number of OB stars populate cavity B \citep{yang1996}, while massive young stellar object candidates align with the shells surrounding cavities A and B \citep{farina2012,martinezgalarza2012}.

Kinematically, these three shells appear to be related. \cite{tenoriotagle2000} found that a single velocity component characterizes the shell around cavity A, while two velocity components characterize the cavity itself. They suggest that these are signs that the bubble originally associated with cavity A had burst in the direction toward us. Meanwhile, the kinematics of cavities B and D were similar to those of cavity A, suggesting a similar origin. 

Meanwhile, the excitation drops in the east (Figure~\ref{excitation}). A prominent loop surrounds cavity C, centered on \mbox{(\ang{;;24}, \ang{;;0})}. The flux intensity here is weaker, and the kinematics do not show line-splitting \citep{tenoriotagle2000}. Only a few stars occupy this cavity; Figure~\ref{pretty_images} marks one of them, an O6.5\,Iaf star \citep{drissen2008,neugent2011}. The kinematics and stellar population here suggest that this fourth cavity is an older structure that has been recently re-ionized by the current burst. 

Two interesting objects lie to the south and will be discussed throughout the paper. One is a previously detected \ac{snr}, M33-16 (also known as L10-124;  \citealt{dodorico1980,long2010,duartepuertas2024,sarbadhicary2024}). The green contour outlines its location and shape as defined in our \sii/\ha\ image (Figure~\ref{excitation}). This gives it an approximate diameter of roughly $\ang{;;4}\times\ang{;;2}$ (\qtyproduct{16 x 8}{\parsec}). Another object to the south of the \ac{snr} is an unknown point source object detected in our \nii/\ha\ and \oiii/\hb\ images (Figure~\ref{excitation}). The green circle is a \ang{;;1} (\SI{4}{\parsec}) aperture. Nearby and marked with a green cross is a potential stellar object detected by \cite{farina2012} with a separation of \ang{;;0.8}.

\subsection{Extinction in NGC~604}

\begin{figure}
\includegraphics[width=\columnwidth,keepaspectratio]{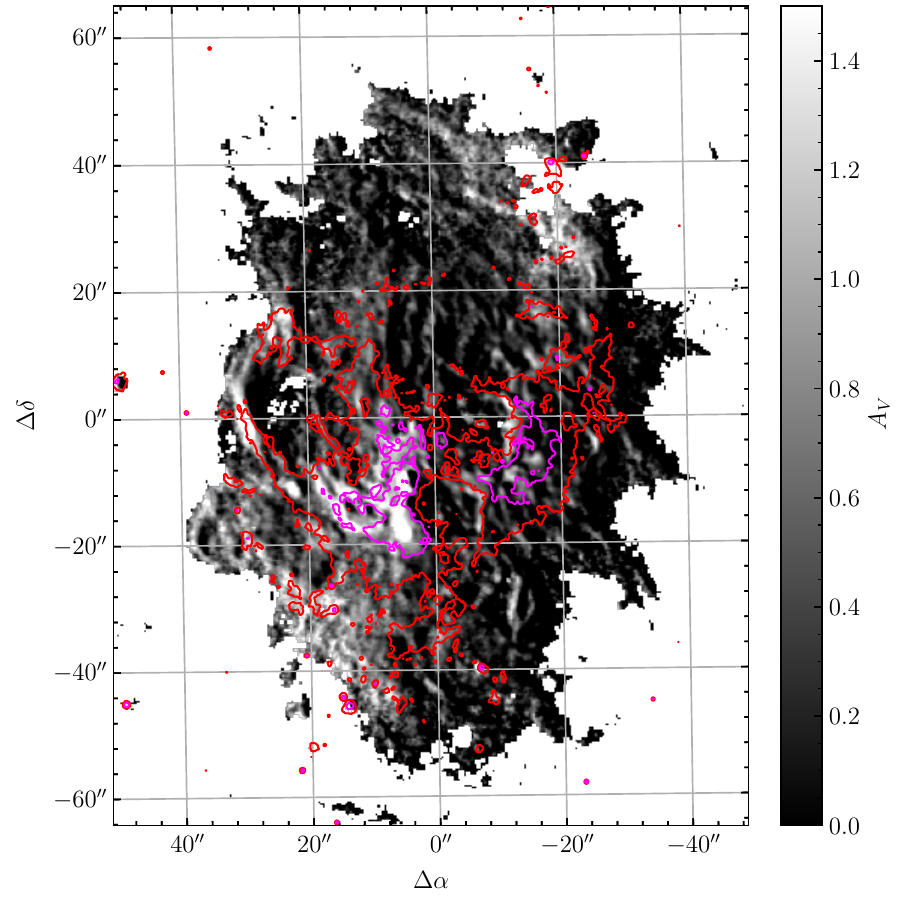}
\caption{The visual dust extinction, $A_V$, map for NGC~604 as estimated from the Balmer decrement. The characteristic median uncertainty is \num{0.17}. The contours show the distribution of the \emph{JWST} \added{MIRI} F770W flux at levels of \qtylist[list-units=single]{20;50}{\mega\jansky\per\steradian} in red and purple, respectively \citep{sarbadhicary2024}.}
\label{Av-image}
\end{figure}

Figure~\ref{Av-image} shows the visual extinction map, $A_V$, derived from the \ha/\hb\ line ratio under the assumption of Case B recombination \citep{osterbrock2006} at $T_e = \SI{8660}{\kelvin}$ \citep{esteban2009} \added{and assuming constant density of \mbox{$n_e = \SI{10}{\per\cubic\centi\metre}$.} Changing the density to values as high as \SI{300}{\per\cubic\centi\metre} does not meaningfully change the intrinsic Balmer decrement at these conditions, namely $(\text{H}\alpha/\text{H}\beta)_{\text{int}} = 2.89$. We adopt the extinction law of \cite{calzetti2000} with $R_V = 3.1$.} As with all subsequent maps, we calculate extinction only for regions where the uncorrected flux in each line exceeds \SI{2e-17}{\erg\per\second\per\square\centi\metre}. The resulting extinction map yields a median value of $A_V = \num{0.22 \pm 0.17}$, slightly lower than, but broadly consistent with, the value derived from the integrated spectrum ($A_V \sim 0.3$). 

The extinction map shows a clumpy and complex spatial structure across NGC~604. Overlaid on Figure~\ref{Av-image} are contours from the \emph{JWST} \added{MIRI} F770W image \citep{sarbadhicary2024}, which trace the \SI{7.7}{\micro\metre} PAH emission. These contours correspond well with regions of elevated extinction, indicating a spatial correlation between the Balmer decrement and infrared dust emission. Previous studies have examined the dust extinction across NGC~604 using various tracers, including the Balmer decrement, infrared emission, and radio emission, and have explored the correspondence between these methods \citep[e.g.,][]{wilson1992,maizapellaniz2004,relano2009,miura2010}. 

These works confirm, for instance, that the prominent extinction peak near the center of the nebula at (\ang{;;6},\ang{;;-15}) corresponds to the main molecular cloud (cloud 2 of \citealt{wilson1992}; cloud 8 of \citealt{engargiola2003,miura2010}), a well-known site of embedded star formation \citep{miura2010}. This cloud also extends eastward and likely contributed to the enhanced extinction in that region \citep{maizapellaniz2004,miura2010}. 

While a detailed spatial analysis of dust extinction and emission is beyond the scope of this paper, it remains an important aspect of understanding the physical conditions in NGC~604. Future work will explore this topic more thoroughly, particularly in the context of how dust geometry, PAH emission, and molecular gas relate to the nebula's ionization and star formation structure.

\subsection{Density Structure}

\begin{figure*}
\includegraphics[width=\textwidth,keepaspectratio]{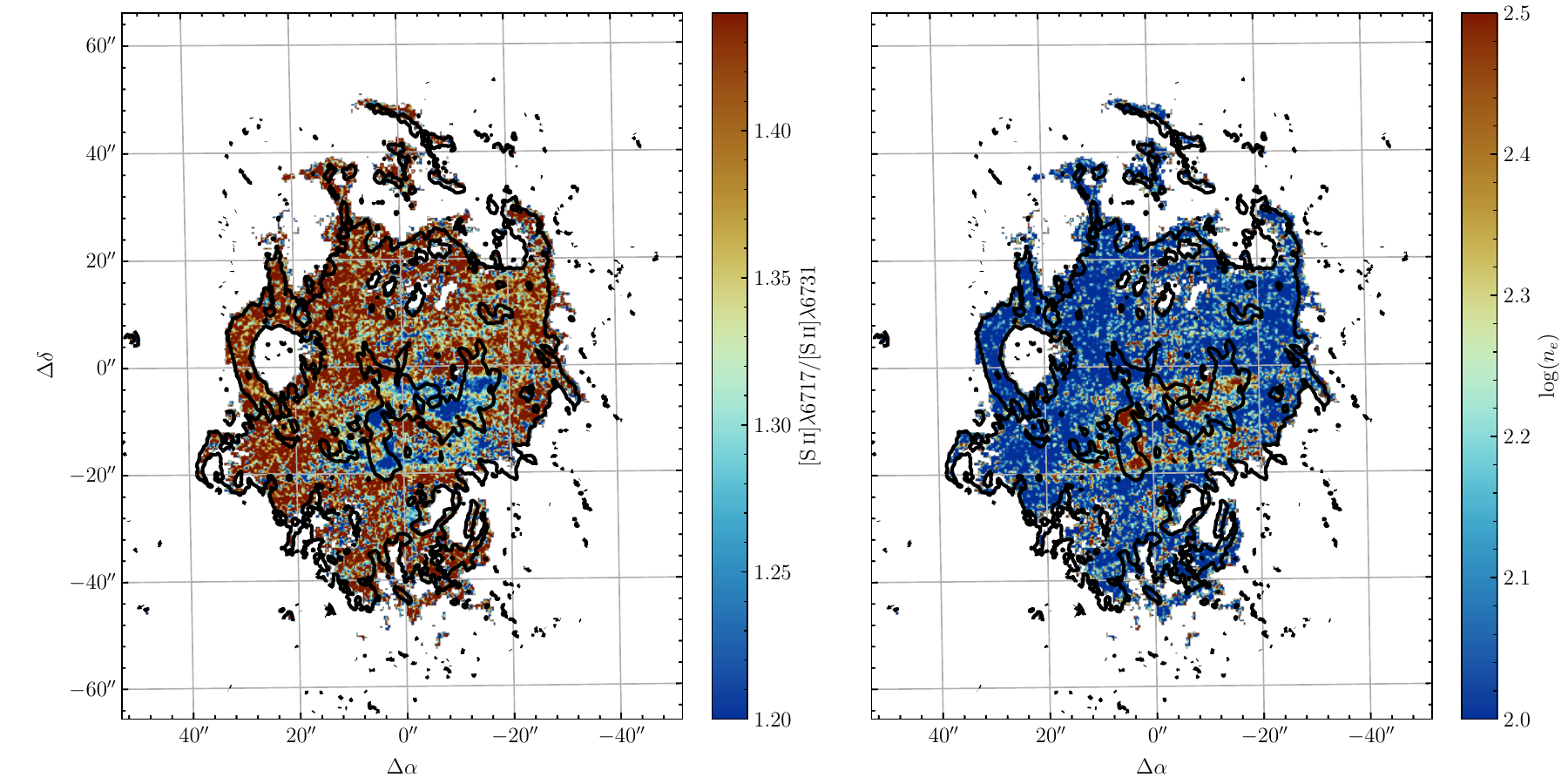}
\caption{Left: the \sii$\lambda$6717/\sii$\lambda$6731 emission line ratio. Characteristic uncertainties are $\sim$0.4. Right: the electron density, $\log n_e$, calculated with \added{\texttt{PyNeb} \citep{luridiana2015}}. Characteristic uncertainties are \SI{1}{\dex}. The black contours show the distribution of the reddening-corrected \ha\ flux at levels of $\log(F_{\text{\ha}}) = -16.5, -15.5, -14.5 \, \si{\erg\per\second\per\square\centi\metre}$. }
\label{s2_density}
\end{figure*}

We study the electron density across NGC~604 using the \sii$\lambda$6717/\sii$\lambda$6731 emission line ratio, shown in the left panel of Figure~\ref{s2_density}. The interquartile range of this ratio is $[1.29,1.49]$, indicating that the region generally lies in the low-density regime, consistent with earlier measurements \citep{vilchez1988,maizapellaniz2004,esteban2009}. The map reveals some notable structures, including clumps in the high-excitation central regions where the ratio is lower than in the surrounding nebula. SNR M33-16, located to the south, also appears on the map with a ratio of approximately $1.3$. 

The right panel of Figure~\ref{s2_density} shows the estimated electron density, $n_e$, \added{calculated with \texttt{PyNeb} \citep{luridiana2015} using the atomic data of \citet[][see \citealt{morisset2020} for a discussion on other atomic data sources]{rynkun2019}. We assume an electron temperature of $T_e = \SI{8680 \pm 200}{\kelvin}$ \citep{esteban2009}. To propagate uncertainties, we employed a Monte Carlo method that incorporates both measurement errors in the emission line ratio and the assumed electron temperature. To speed up computation, those pixels with the \sii$\lambda$6717/\sii$\lambda$6731 ratio greater than $1.35$, we assigned an electron density of \SI{100}{\per\cubic\centi\metre}. For all pixels, the resulting $\log n_e$ values typically have uncertainties on the order of \SI{1}{\dex}.}

Most of the \hii\ region lies in the low-density limit, with $n_e < \SI{100}{\per\cubic\centi\metre}$. However, several denser clumps are apparent, particularly around the bright central high-excitation region, where $n_e \gtrsim\SI{300}{\per\cubic\centi\metre}$. The densest knot in NGC~604 is located at (\ang{;;8},\ang{;;-8}), with a peak value near \SI{500}{\per\cubic\centi\metre}. This location also corresponds to a region of high extinction in the $A_V$ map (Figure~\ref{Av-image}). While the density is consistent with a compact \hii\ region, the knot's size of about \SI{5}{\parsec} in the \ha\ map is an order of magnitude too large to qualify as a classical compact \hii\ region (see \citealt{kurtz2002} for a review). Instead, it may represent a filled \hii\ region ionized by a small group of massive stars. This same region also exhibits the highest excitation in NGC~604 with $\log(\text{\oiii}/\text{\hb}) \simeq 0.7$ (Figure~\ref{excitation}). 

Other areas of enhanced density include the bright arc that outlines cavity A. \cite{maizapellaniz2004} proposed that this region consists of a dusty ``flap'' that obscures part of the near-edge-on high-excitation zone. In this scenario, we may be observing line emission from one surface of the cavity passing through the dust flap, artificially elevating the measured density due to projection effects. 

Finally, SNR M33-16, though not as dense as the previous regions, is modestly enhanced relative to the surrounding \ac{ism}, with densities around \SI{145}{\per\cubic\centi\metre}. This density is notably lower than the value of $\sim$\SI{445}{\per\cubic\centi\metre} reported by \cite{gordon1998}, but given the large uncertainties on our electron density estimates of $\pm\SI{1}{\dex}$, this difference is not cause for alarm.

\subsection{Excitation Maps}\label{sec:excitation_maps}

\begin{figure*}
\includegraphics[width=\textwidth,keepaspectratio]{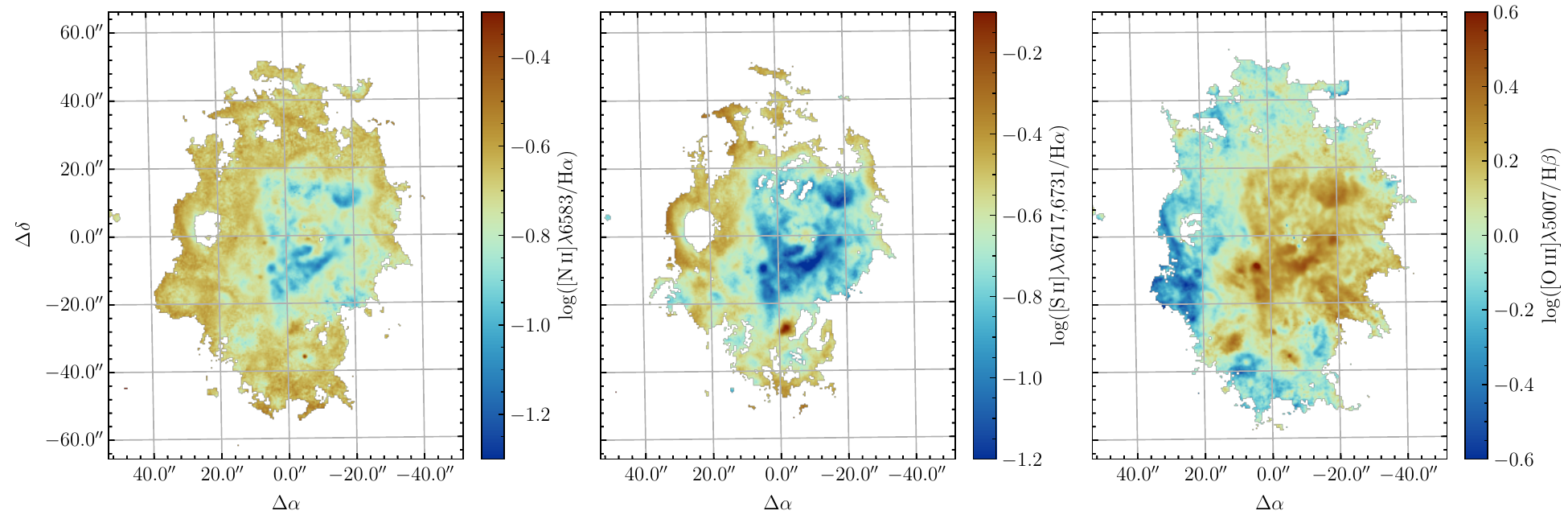}
\caption{From left to right: the $\log(\text{\nii}\lambda6584/\text{\ha})$, $\log(\text{\sii}\lambda\lambda6717,6731/\text{\ha})$, and $\log(\text{\oiii}\lambda5007/\text{\hb})$ emission line ratio maps. Characteristic uncertainties are \qtylist{0.08;0.08;0.11}{\dex}, respectively. North is up, east to the left.}
\label{excitation}
\end{figure*}

Figure~\ref{excitation} shows the emission line ratio maps of \nii/\ha, \sii/\ha, and \oiii/\hb. These ratios allow us to study the ionization structure of NGC~604, with the former two ratios tracing the low-excitation regions and the latter tracing the high-excitation regions. Lower values of \nii/\ha\ and \sii/\ha\ are located close to the position of the main ionizing star cluster, corresponding to the high-excitation region. In comparison, the outskirts have higher values, indicating the low-excitation zone. The opposite is true for the \oiii/\hb\ map. 

We immediately see the striking straight line ``ridge'' remarked on by \cite{tenoriotagle2000}: a large, almost straight dividing line that spans across the center of the nebula from north to south. This ridge separates the high-excitation region from the low-excitation region to the east. They suggest that three shells or bubbles happen to align by chance, and because we view them edge-on, they appear as a straight-line ridge. Interestingly, the \nii/\ha\ and \sii/\ha\ ratios somewhat maintain this sharp dividing line, while the \oiii/\hb\ ratio appears to have breached the ridge as there are patches of high-excitation regions towards the east. However, this appears to be the location of a few \ac{wr} stars that might be ionizing the surrounding \ac{ism} \citep{bruhweiler2003,neugent2011}. Except for these few regions, the eastern section of the nebula seems unaffected by the ionization occurring in the central cluster. For instance, the prominent arc that forms the boundary of cavity C is consistent with very low excitation, which supports an old age for this portion of the nebula \citep{tenoriotagle2000,maizapellaniz2004}. 

\begin{figure*}
\includegraphics[width=\textwidth,keepaspectratio]{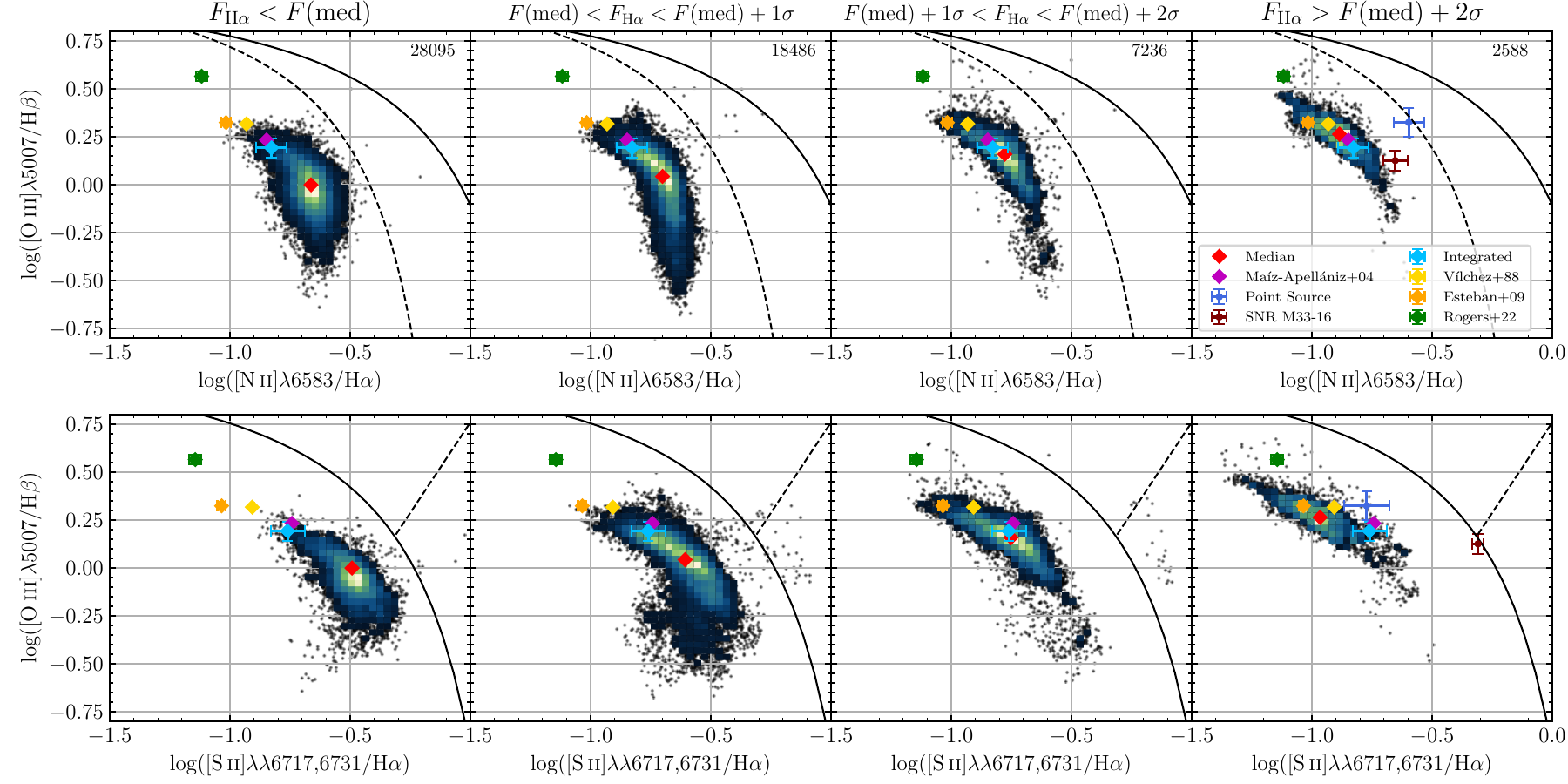}
\caption{The \nii\ (top panels) and \sii\ (bottom panels) BPT diagrms \citep{baldwin1981,veilleux1987} for all of the pixels in the nebula where we measure \oiii/\hb, \nii/\ha, and \sii/\ha\ in bins of different \ha\ fluxes as indicated by the panel titles. The numbers in the top right corner of the top row of panels indicates how many pixels are in each panel. The diamonds indicate observed emission line ratios in this paper and from the literature \citep{vilchez1988,maizapellaniz2004,esteban2009,rogers2022} indicated in the legend; the blue diamond is the integrated value reported in Section~\ref{sec:integrated} while the red diamond is the median values in each panel. \added{The two circles in the rightmost plot show the location of the point source and \ac{snr} M33-16 as discussed in Section~\ref{sec:discrete}.} In the top panels, the dashed line is the pure star formation demarcation from \cite{kauffmann2003}. In all panels, the solid line is the extreme starburst demarcation from \cite{kewley2001}. In the bottom panels, the dashed line is the Seyfert/LINER line from \cite{kewley2006}.}
\label{bpt}
\end{figure*}

Using BPT diagrams \citep{baldwin1981,veilleux1987}, we investigate whether other non-photoionizing mechanisms are significant with NGC~604. In Figure~\ref{bpt}, we show the BPT diagrams grouped into bins of \ha\ flux. Those pixels with high \ha\ flux (right-most panels) correspond to those showing high excitation levels (high values of \oiii/\hb); however, the low \ha\ flux pixels (left-most panels) cover the whole range of excitation within the nebula. Interestingly, the range covered by all bins in \oiii/\hb\ is approximately the same. However, as indicated by the position of the red diamonds in Figure~\ref{bpt}, which indicate our median values in each flux bin, higher \nii/\ha\ and \sii/\ha\ ratios and lower \oiii/\hb\ ratios are detected in areas of lower \ha\ flux. This quantitative trend confirms the qualitative trend seen in Figure~\ref{excitation} that the degree of ionization does get smaller with increasing distance from the ionizing source. 

\added{A recent analysis of 124 \hii\ regions in the southern part of M33 by \cite{feltre2025} showed similar results. Lower values of \nii/\ha\ and \sii/\ha\ occur where the ionization is strongest (as traced by \siii$\lambda\lambda$9069,9531/\sii$\lambda\lambda$6717,6731) and vice versa, at least out to the ionization front where these values plateau. Meanwhile, the \oiii/\hb\ ratio appears to remain approximately constant likely due to gas physical conditions and the optical depth. This is evident in our BPT diagrams, where we find that while \nii/\ha\ and \sii/\ha\ strongly vary across our flux bins, \oiii/\hb\ has weaker variations. \cite{feltre2025} proposed that this plateau of \oiii\ emission is caused by complex nebular geometry which allows photons to leak along multiple sightlines of low optical depth (see also \citealt{pellegrini2012,jin2022}). The complex geometry of NGC~604 (Figure~\ref{pretty_images}) could be the culprit here as well. Indeed, \cite{maizapellaniz2004} found that it is density-bounded to the west, resulting in one direction where photons might be leaking and maintaining high \oiii/\hb\ ratios.}

In general, these findings compare well with previous studies of two other spatially-resolved \hii\ regions in M33, NGC~595, and NGC~588 \citep{relano2010,monrealibero2011}. In general, the trends observed across the changing brightness bins are the same, although the ranges of these line ratios are smaller in NGC~595 and NGC~588. As \cite{monrealibero2011} remarks, this is likely due to different mapped areas. As both are within the \ac{signals} data for M33, a comparison study between our data would be interesting. 

%We also show the separation between active galactic nuclei and normal star-forming galaxies from different studies \citep{kewley2001,kauffmann2003,kewley2006}. 
%
%
%As can be seen in the figure, most of the pixels lie below the separation lines indicating that the main mechanism for producing the emission lines is photoionization by OB stars. A few pixels cross these separation lines; those in the \sii\ BPT diagram (bottom row of Figure~\ref{bpt}) are part of the known \ac{snr} M33-16, while those in the \nii\ BPT diagram (top row of Figure~\ref{bpt}) is a point-source object also visible in the left panel of Figure~\ref{excitation} towards the south. Despite these localized regions, most of NGC~604 is dominated by photoionization from OB stars. 

Also marked in Figure~\ref{bpt} are the observed integrated emission line ratios from this study (blue diamond) and others \citep{vilchez1988,maizapellaniz2004,esteban2009,rogers2022}. It appears that, independently of the line ratio, integrated values from long-slit spectroscopy (yellow, orange, green diamonds) are more representative of the high surface brightness areas (right-most panels). In contrast, those from imaging or imaging spectroscopy (blue and purple diamonds) are more representative of middling surface brightness ($F(\text{med}) + 1\sigma < F_{\text{\ha}} < F(\text{med}) + 2\sigma$). Moreover, the \nii/\ha\ and \sii/\ha\ ratios derived from long-slit spectroscopy  differ by as much as $\sim$\SI{0.3}{\dex} and $\sim$\SI{0.4}{\dex}, respectively, from those derived from imaging or imaging spectroscopy. Similar effects have been observed in other giant \hii\ regions \citep{pellegrini2010,relano2010,monrealibero2011}. These trends illustrate the effect of aperture selection and relative weighting of external regions of an \hii\ region, as imaging and imaging spectroscopy will naturally weigh external, low excitation regions more, while long-slit spectroscopy often give more weight to high-excitation regions \added{as is required to measure the faint \oiii$\lambda$4363 line. }

\subsubsection{Discrete Sources}\label{sec:discrete}

Two discrete sources are evident in Figure~\ref{excitation}, which warrant further discussion. In the \sii/\ha\ excitation map (middle panel), the \ac{snr} M33-16 at (\ang{;;-2},\ang{;;-28}) stands out at logarithmic values between \qtyrange[range-units=single]{-0.3}{-0.1}{\dex}, consistent with \acp{snr} \citep{dodorico1976,yang1996,long2010}. For completeness, we measured the integrated flux of the \ac{snr} in a custom contour defined using the \sii/\ha\ line ratio map and shown in Figure~\ref{pretty_images}. The observed spectrum did not show any signs of line-splitting or Doppler broadening consistent with being an old \ac{snr} \citep{long2018}. Therefore, we only assumed a single velocity component in fitting the spectrum and deriving line fluxes. \added{We fit a phase-shifted sincgauss profile in SN3 and a standard sincgauss profile in SN1 and SN2 with \textsc{orcs}.} We did not correct the fluxes for the complex \hii\ region background, but we did apply standard dust corrections using the Balmer decrement and the extinction law of \cite{calzetti2000}. Those emission line fluxes and selected line ratios are reported in Table~\ref{odd-objects}. \added{Given its integrated \ha\ brightness, we show its location in the BPT diagram in the far right plots of Figure~\ref{bpt} as the maroon point.}

\begin{deluxetable}{l c c}
\tablecaption{Integrated Line Fluxes for Two Objects \label{odd-objects}}
\tablehead{\colhead{Line/Line Ratio} & \colhead{SNR M33-16} &\colhead{Point Source}}
\startdata
\oii$\lambda$3727 & \num{5.034 \pm 0.702} & \num{3.192 \pm 0.632}   \\
\hb\ $\lambda$4861 & \num{1.000 \pm 0.117} & \num{1.000 \pm 0.167}   \\
\oiii$\lambda$4959 & \num{0.466 \pm 0.055} & \num{0.735 \pm 0.123}  \\
\oiii$\lambda$5007 & \num{1.337 \pm 0.154} & \num{2.114 \pm 0.344} \\
\nii$\lambda$6548 & \num{0.259 \pm 0.058} & \num{0.358 \pm 0.086} \\
\ha\ $\lambda$6563 & \num{2.859 \pm 0.291} & \num{2.860 \pm 0.409}  \\
\nii$\lambda$6584 & \num{0.641 \pm 0.082} & \num{1.084 \pm 0.172} \\
\sii$\lambda$6717 & \num{0.792 \pm 0.095} & \num{0.294 \pm 0.081}   \\
\sii$\lambda$6731 & \num{0.614 \pm 0.079} & \num{0.216 \pm 0.076}   \\
$F(\mathrm{H}\beta) \, \lambda 4861$ & \num{3.341 \pm 0.277} & \num{0.283 \pm 0.033}   \\
$E(B-V)$ & \num{0.131 \pm 0.024} & \num{0.001 \pm 0.034}   \\
& & \\
$\log\text{\nii}\lambda6584/\text{\ha}$ & \num{-0.649 \pm 0.049} & \num{-0.421 \pm 0.058}   \\
$\log\text{\sii}\lambda\lambda\text{6717,6731}/\text{\ha}$ & \num{-0.308 \pm 0.038} & \num{-0.749 \pm 0.094}   \\
$\log\text{\oiii}\lambda5007/\text{\hb}$ & \num{0.126 \pm 0.050} & \num{0.325 \pm 0.071}   \\
$\log\text{\oii}\lambda3727/\text{\hb}$ & \num{0.702 \pm 0.061} & \num{0.504 \pm 0.086}   \\
$\log R_{23}$ & \num{0.835 \pm 0.051} & \num{0.781 \pm 0.066}  \\
$\log \text{O}_{32}$ & \num{-0.446 \pm 0.056} & \num{-0.049 \pm 0.079}   \\
$\log\text{\nii}\lambda6584/\text{\oii}\lambda3727$ & \num{-0.895 \pm 0.064} & \num{-0.469 \pm 0.083}   \\
\sii$\lambda$6717/\sii$\lambda$6731 & \num{1.290 \pm 0.168} & \num{1.361 \pm 0.565}   \\
\enddata
\tablecomments{The first column corresponds to the emission line or emission line ratio identification. The fluxes are all reddening-corrected and normalized to the reddening-corrected \hb\ flux. The \hb\ fluxes are in units of \SI{e-14}{\erg\per\second\per\square\centi\metre}.}
\end{deluxetable}

Another discrete source lies south of \ac{snr} M33-16 at (\ang{;;-5},\ang{;;-36}) and is clearly visible in the \nii/\ha\ and \oiii/\hb\ maps in Figure~\ref{excitation}. This is consistent with an object that is locally nitrogen-enriched and has a high-excitation source. Given its small point-like size ($r \sim \ang{;;1}$; $\sim$\SI{4}{\parsec}) and its location on the BPT diagrams, this suggests that this object is a planetary nebula. We extracted its spectrum from the data cube using a \ang{;;1} aperture and derived its line fluxes. In order to somewhat account for the \hii\ region background, we performed local background subtraction using an annulus with inner radius \ang{;;2.3} and width \ang{;;1.3}. However, upon subtracting this local background, only the \nii\ and \oiii\ emission lines remained, indicating that this object has excess \nii\ and \oiii, perhaps in agreement with observations of Type I planetary nebulae \citep{kingsburgh1994,frew2010}. \added{Unfortunately, no other emission lines remain after local background subtraction, precluding further analysis.}

Additionally, before we subtracted the local background, we noticed that the \nii\ lines lie at a different velocity than that of NGC~604 (\SI{-255}{\kilo\metre\per\second}; \citealt{tenoriotagle2000}). Namely, this point source is redshifted by about \SI{80}{\kilo\metre\per\second}. This suggests that this point source does not belong to NGC~604 from a kinematic perspective. Putting this together suggests that this point source is a rare Type I planetary nebula that we are observing along the line-of-sight towards NGC~604. In Table~\ref{odd-objects}, we include the emission line fluxes for this source and selected line ratios. \added{We fit two velocity components with a phase-shifted sincgauss profile in SN3 and a single-component standard sincgauss profile in SN1 and SN2 with \textsc{orcs}.} Similar to \ac{snr} M33-16, the fluxes in this table do not include local background subtraction but do include the standard dust correction assuming the \cite{calzetti2000} extinction law. \added{We also show its location in the BPT diagram in Figure~\ref{bpt}. Given its \ha\ brightness, it is located in the far right plots of Figure~\ref{bpt} as the dark blue circle.}

%Despite this similarity, applying a common planetary nebula diagnostic \citep{ciardullo2002} shows that it is not a planetary nebula. 
%
%However, before we subtracted the local background, we noticed that the \nii\ lines show evidence of line-splitting (Figure~\ref{weird_spec}). We fit this un-corrected spectrum with two velocity components and found a component redshifted by $\sim$\SI{80}{\kilo\metre\per\second} relative to the velocity of NGC~604 (\SI{-255}{\kilo\metre\per\second}; \citealt{tenoriotagle2000}). This might indicate some sort of wind, perhaps from a \ac{wr} star, but no other spectral features indicative of a \ac{wr} star, such as broad lines or helium lines \citep{crowther2007}, are seen. 

Finally, there are four additional point sources in the \nii/\ha\ map, albeit fainter than the strong point source to the south. Three of them, at coordinates (\ang{;;-7},\ang{;;-1}), (\ang{;;6},\ang{;;-2}), and (\ang{;;8},\ang{;;-12}), are known WN stars \citep{drissen2008}. The fourth at (\ang{;;15},\ang{;;-1}) is not in the catalog of \cite{drissen2008}, but is a massive young stellar object in the catalog of \cite{farina2012}. WN stars do not exhibit \nii\ lines, unless they are surrounded by a wind-driven bubble enriched in nitrogen which might be the case for these four stars.

\subsection{Strong-line Ratio Maps}\label{sub:strong_ratio_maps}

\begin{figure*}
\includegraphics[width=\textwidth,keepaspectratio]{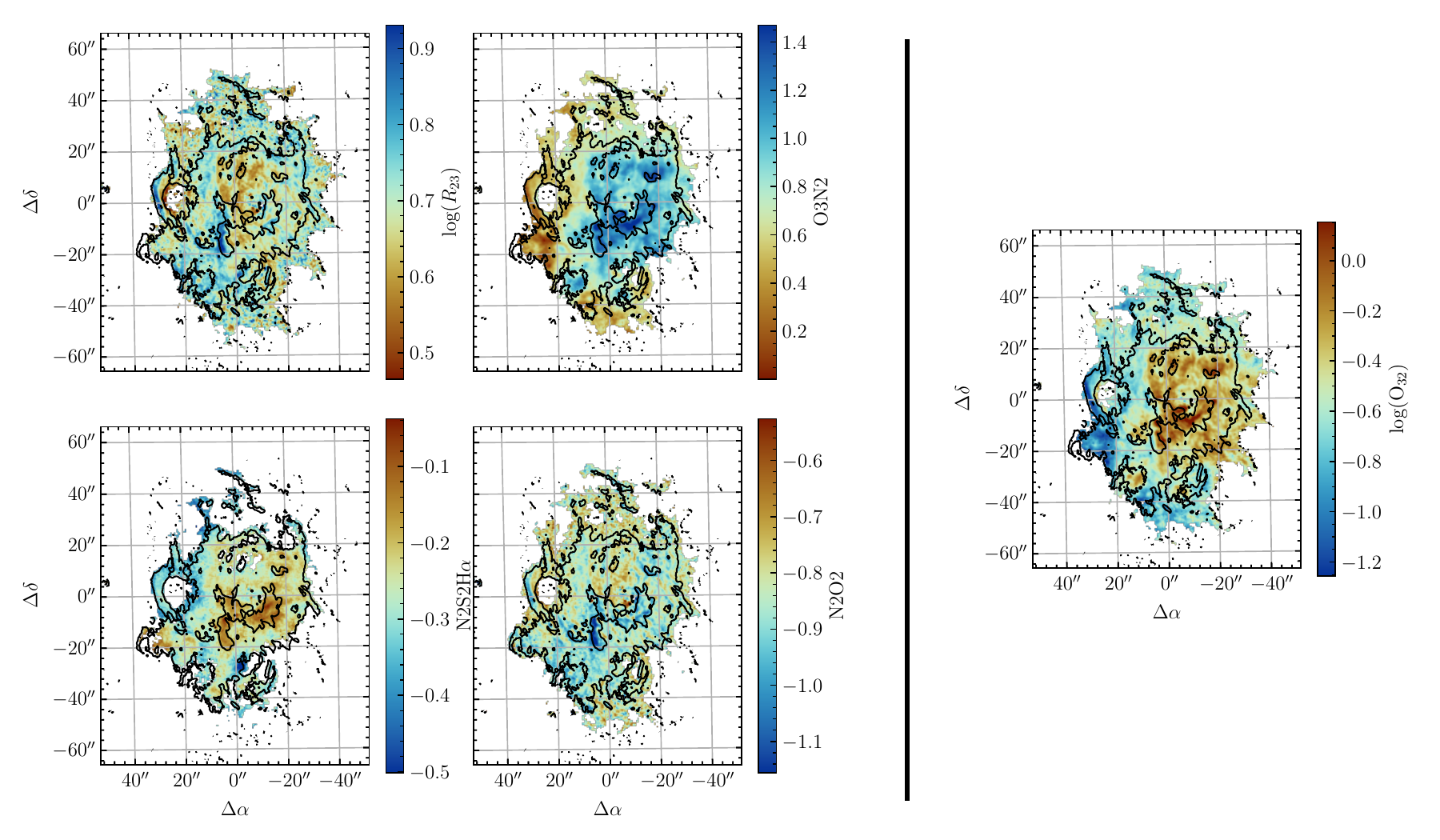}
\caption{Maps of common strong-line abundance and ionization indicators. The left set of four panels are common abundance indicators: the $R_{23}$ map (top left; \citealt{pagel1979}), the O3N2 map (top right; \citealt{marino2013}), the N2S2\ha\ map (bottom left; \citealt{dopita2016}), and the N2O2 map (bottom right; \citealt{kewley2019}). Each color bar has been chosen so that bluer colors are metal-poor while redder colors are metal-rich. Characteristic uncertainties on each are \qtylist[list-units=single]{0.13;0.14;0.10;0.13}{\dex}, respectively. The right plot is the ionization-sensitive O$_{32}$ map \citep{kewley2002}. In this case, bluer colors are less ionized while redder colors are more ionized. The characteristic uncertainty on this map is \SI{0.14}{\dex}. In all plots, the range on each color bar is $\pm 3\sigma$ around the median and the black contours show the distribution of the reddening-corrected \ha\ flux at levels of $\log(F_{\text{\ha}}) = -16.5, -15.5, -14.5 \, \si{\erg\per\second\per\square\centi\metre}$.}
\label{abund_ratios}
\end{figure*}

Our imaging spectroscopy also enables us to investigate the distribution of line ratios commonly used to estimate the metallicity and ionization parameter of integrated \hii\ regions throughout NGC~604 for the first time. Namely, we use all of the abundance- and ionization-sensitive diagnostics defined in Section~\ref{sub:oxy}, namely the $R_{23}$, O$_{32}$, O3N2, N2S2\ha, and N2O2 line ratios (Equations~\ref{eq:r23}-\ref{eq:n2o2}).

Figure~\ref{abund_ratios} shows the maps for these line ratios. The right panel shows the ionization-sensitive indicator, O$_{32}$. There is a clear trend of decreasing O$_{32}$ with radial distance from the ionizing stars. This follows from the definition of the ionization parameter \citep{kewley2002,kewley2019}. Again, there is a definite morphological structure in the O$_{32}$ map, just as there was in the excitation maps. We see the ``ridge'' dividing the less-ionized east from the highly ionized west. The interquartile range for this parameter is $[-0.69, -0.39]$. The highest O$_{32}$ values are located in the compact \hii\ region near the center with values reaching $\log\text{O}_{32} \simeq 0.8$. The point source object to the south also shows mildly high values of $\log\text{O}_{32} \simeq -0.1$ (Table~\ref{odd-objects}), indicating a mildly ionizing source. 

Meanwhile, the left group of panels in Figure~\ref{abund_ratios} shows the abundance-sensitive indicators. Looking first at the $R_{23}$ map (top left), it shows very little variation with morphology. The interquartile range for this parameter is $[0.65, 0.74]$, comparable to our integrated value (Table~\ref{int-ratios}). Despite the $\pm$\SI{0.05}{\dex} variation away from our integrated value, this is within the characteristic uncertainty for this parameter of \SI{0.13}{\dex}. Thus, it appears that the $R_{23}$ has a relatively constant value across an \hii\ region despite strong variations in the ionization parameter \citep{kennicutt2000,oey2000,james2016,mao2018}\footnote{While the $R_{23}$ famously has a dependence on the ionization parameter, on the upper metal-rich branch of the $R_{23}$-O/H relation, the dependence on ionization is small \citep[e.g.,][]{kobulnicky2004}.}. 

The panels of Figure~\ref{abund_ratios} showing the O3N2 (top right), N2S2\ha\ (bottom left), and N2O2 (bottom right) maps are commonly used as oxygen abundance indicators, but each has secondary dependencies: all of them are dependent on nitrogen abundance, while the O3N2 ratio is additionally dependent on the ionization parameter. The O3N2 map has a superficial similarity to the O$_{32}$ map with strong variations across the nebula. Meanwhile, the N2O2 map has a superficial similarity to the $R_{23}$ map with relatively constant values across the entire region; the median value is \SI{-0.85}{\dex}. Finally, the N2S2\ha\ map exhibits mild fluctuations across the nebula. Notably, its use of the \sii\ lines allows an easy identification of \ac{snr} M33-16. For reference, the interquartile ranges of these parameters are
\begin{align*}
	\text{O3N2} &\in [0.59, 0.92], \\
	\text{N2S2\ha} &\in [-0.32, -0.22], \text{ and} \\
	\text{N2O2} &\in [-0.90, -0.78].
\end{align*}

The natural next step is to apply the various oxygen abundance and ionization calibrations to this data, thus producing oxygen abundance and ionization parameter maps for NGC~604. However, using calibrations on sub-\hii\ region-scale data is not applicable \added{(see \citealt{stasinska2019} for a relevant discussion)}. Empirical calibrations are calibrated on integrated spectra of \hii\ regions or even whole galaxies \citep[e.g.,][]{pettini2004,pilyugin2005,marino2013}. In contrast, theoretical calibrations assume a central ionizing stellar population \citep[e.g.,][]{mcgaugh1991,kobulnicky2004,dopita2016,kewley2019} and it is extremely unlikely that there is an ionizing star in each pixel of our data. Therefore, we construct ``residual'' maps of these strong-line ratios by subtracting the integrated spectrum line ratio from the maps. Thus, we can investigate fluctuations across NGC~604 without quantifying a specific abundance or ionization parameter, both of which are inherently calibration-specific and only applicable to integrated \hii\ regions. 

%%% What do these residuals mean in terms of Delta O/H? Make a figure, but you don't have to use it
%%% How badly do these calibrations behave? 

%%% M91 R23: pm0.3
%%% KK04 R23: pm0.3
%%% KK04 O32: pm0.6
%%% PT05 R23: pm0.3
%%% M13 O3N2: pm0.15
%%% D16 N2S2Ha: pm0.2
%%% K19 O32: pm0.5
%%% K19 N2O2: pm0.1

\begin{figure*}
\includegraphics[width=\textwidth,keepaspectratio]{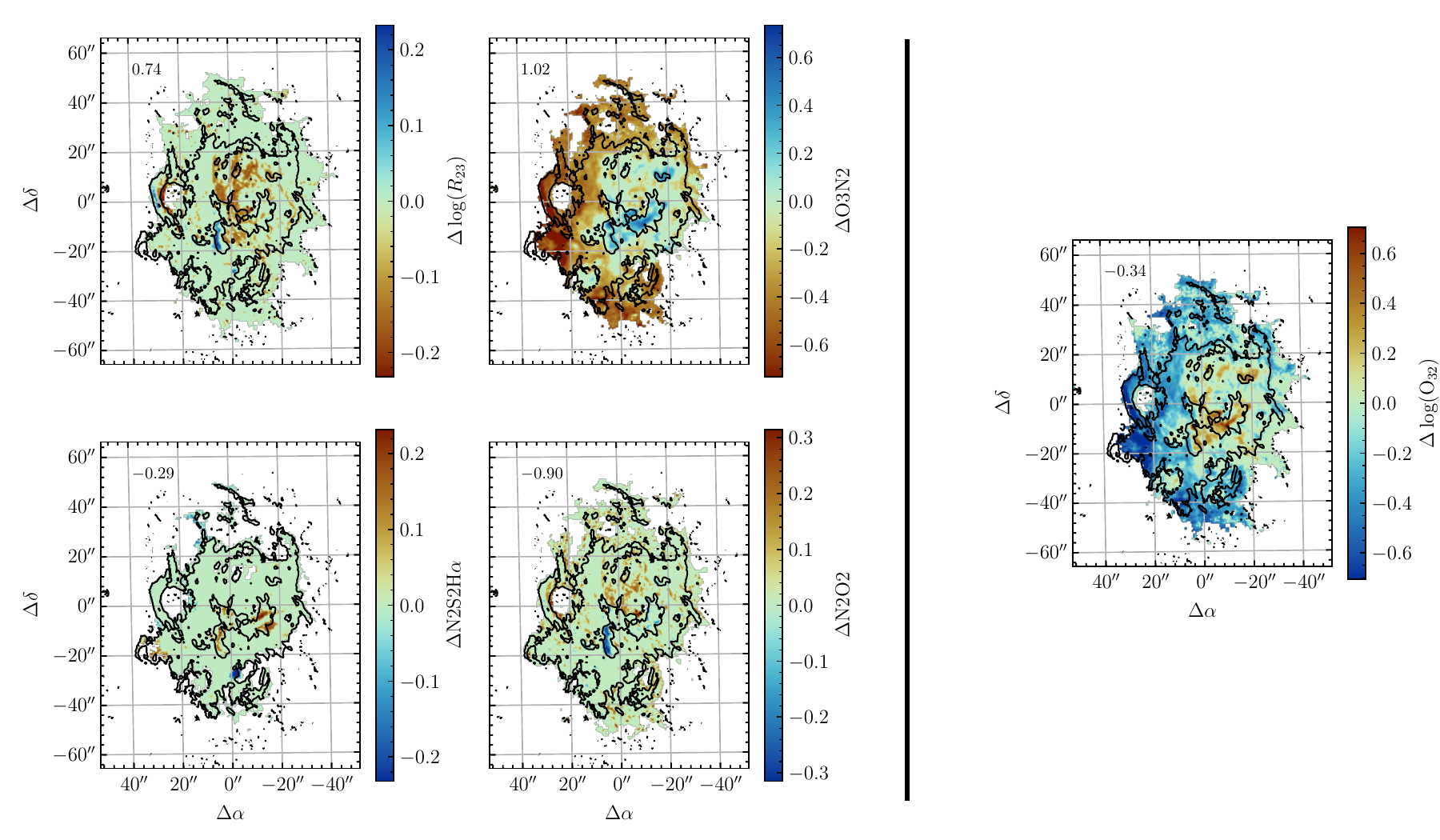}
\caption{Maps of the strong-line abundance indicators after subtracting the value measured in the integrated spectrum. The maps are in the same layout as Figure~\ref{abund_ratios}. The number in the top left of each panel is the line ratio value of the integrated spectrum. The color bars have been chosen so that bluer colors are metal-poor compared to the integrated spectrum while redder colors are metal-rich compared to the integrated spectrum, while light green are pixels near a value of zero that have been masked (see the text). The range on each color bar is $\pm 3\sigma$ around zero. Characteristic uncertainties on each are \qtylist[list-units=single]{0.14;0.15;0.16;0.13;0.15}{\dex}. The black contours show the distribution of the reddening-corrected \ha\ flux at levels of $\log(F_{\text{\ha}}) = -16.5, -15.5, -14.5 \, \si{\erg\per\second\per\square\centi\metre}$. }
\label{residual_maps}
\end{figure*}

However, not all fluctuations of these line ratios around the integrated value are statistically significant. In many cases, the residuals are smaller than the uncertainties on the residuals. In that case, we choose instead to mask these areas of NGC~604, as seen in Figure~\ref{residual_maps}, where we have set these pixels to have a residual of zero and colored them light green. We selected all other colors so that, in the case of an integrated \hii\ region given those line ratios, red is metal-rich while blue is metal-poor. This color choice required flipping the color bars on the $R_{23}$ and O3N2 maps where more positive values correspond to lower metallicity \citep{mcgaugh1991,kobulnicky2004,marino2013}. In the case of the O$_{32}$ map, more positive values indicate higher ionization and are colored red, while more negative values represent lower ionization and are colored blue in the case of an integrated \hii\ region. Each color bar spans a range of $\pm 3\sigma$. 

Through this lens, we observe that some line ratios exhibit significant variations while others do not. The residuals in the line ratios, which show little dependence on the ionization parameter, namely $R_{23}$, N2S2\ha, and N2O2, exhibit very weak variations consistent with zero change from the integrated value. Thus, most of NGC~604 in these maps are colored light green. However, a few statistically significant features remain. For instance, in the $R_{23}$ and N2O2 maps, a ``three-pronged fan'' to the north appears to have consistently higher line ratios relative to the integrated value. Given the concentric nature of these arcs, they may represent remnants of shocks or the outer edges of bubbles from early star formation episodes. Indeed, shocks are known to change the value of the $R_{23}$ index \citep[e.g.,][]{dopita1996}. 

Other smaller discrete features are also visible in these abundance-sensitive line ratio maps. The large arc to the east appears to be bifurcated, with the western side having higher values than the eastern side. This is more easily seen in the $R_{23}$ map than in the N2O2 map. \cite{tenoriotagle2000} regarded this structure as old since it shows no evidence of line-splitting and is thus likely unaffected by the mechanical energy from massive stars that the central region feels. However, there is at least one massive star near this arc \citep{drissen2008} on the western side, so metal enrichment by local star formation or their shocks is not ruled out regardless of the structure's supposed old age. 

In both the $R_{23}$ and N2O2 residual maps, the site of the main molecular cloud at (\ang{;;6},\ang{;;-15}) has lower values compared to the integrated value, while in the N2S2\ha\ residual map it is has higher values. Importantly, this area has the highest dust extinction of the region (Figure~\ref{Av-image}). This dust-enshrouded area does have active star formation within it. However, the dust is likely shielding its light, especially so close to the central ionizing cluster. Other spots in the N2S2\ha\ residual map do not appear in any other map, SNR M33-16 being the most prominent example as having much lower values compared to the integrated value for the entire region. Comparing this residual map to the electron density maps in Figure~\ref{s2_density}, we see that the areas here that are not masked are where we have reliable density measurements. There is likely some electron density dependence not being controlled for in this abundance diagnostic, especially since supernovae enrich the local \ac{ism} with metals \citep[e.g.,][]{timmes1995}. 

Another factor to consider is the ionization parameter, which likely explains the large variations observed in the O$_{32}$ and O3N2 residual maps. Among the lines used in these line ratios, \oiii\ is more efficiently excited in regions of high ionization, while \oii, \nii, and \sii\ are more related with a low-ionization state \citep{osterbrock2006}. Decreasing the ionization parameter would have the effect of diminishing O3N2 and O$_{32}$ and to keeping $R_{23}$, N2O2, and N2S2\ha\ constant if intrinsic metallicities are constant. Indeed, photoionization models configured with such an ionization structure have successfully reproduced the observed 2D features of other individual \hii\ regions in M33 \citep{perezmontero2011,perezmontero2014}. The variation we see in the O3N2 and O$_{32}$ maps in Figures~\ref{abund_ratios}~and~\ref{residual_maps} is likely a result of this changing ionization parameter across NGC~604. 

The above discussion indicates that, despite the complex morphological and structural properties of NGC~604, the abundance-sensitive line ratios are approximately constant across the region, while the ionization structure varies. Although these indices have other dependencies, their approximately constant values demonstrate robustness against ionization variations, a conclusion supported by both observations \citep{kennicutt2000,oey2000,bresolin2007,james2016} and theoretical predictions \citep{kewley2002,dopita2013}. Meanwhile, the failure of the O3N2 index to account for changing ionization states results in a blend of abundance and ionization effects that render the use of this index difficult, especially for resolved studies of \hii\ regions \citep{stasinska2010,lopezsanchez2010,marino2013,mao2018}. Finally, while the N2S2\ha\ index is not sensitive to the ionization state, its dependence on the N/O and S/O relative abundances, as well as  possibly the electron density, hampers its use without prior knowledge of the relative abundances.

\section{Kinematics of NGC~604}\label{sec:kinematics}

The kinematics of NGC~604 has been studied numerous times in the literature \citep[e.g.,][]{smith1970,melnick1977,melnick1980,dodorico1981,rosa1982,rosa1984,hippelein1984,sabalisck1995,yang1996,medinatanco1997,tenoriotagle2000}. While the spectral resolution in our SN3 filter allows us to derive the \ha\ radial velocity and velocity dispersion across the entire nebula, the resolution is not high enough to make a study of split line profiles and thus turbulence. Instead, we offer a ``holistic'' view of NGC~604's kinematics across the entire nebula and make connections to the ionized gas properties discussed above. 

\begin{figure*}
\includegraphics[width=\textwidth,keepaspectratio]{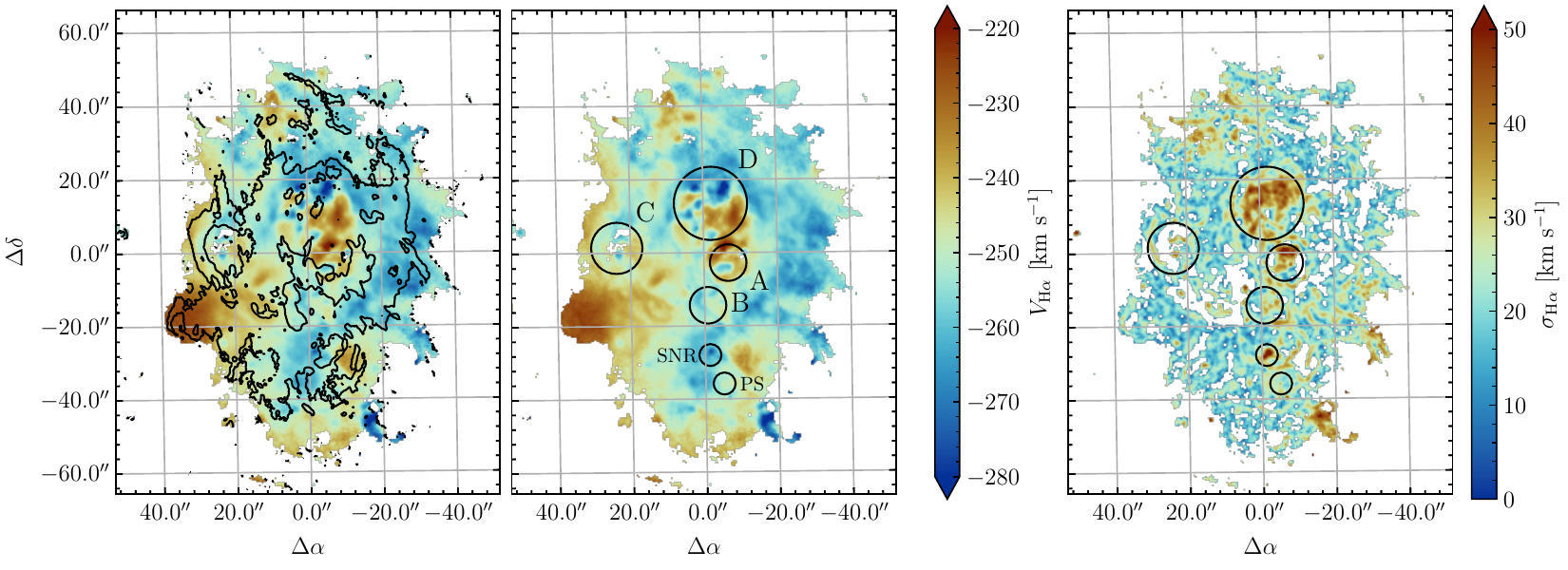}
\caption{Left and middle: the velocity map derived from the \ha\ emission line. Overlaid in black on the left panel are \ha\ flux contours at levels of $\log(F_{\text{\ha}}) = -16.5, -15.5, -14.5 \, \si{\erg\per\second\per\square\centi\metre}$. In the middle panel, we have marked the positions of the four main cavities A through D from \cite{maizapellaniz2004} as well as \ac{snr} M33-16 (SNR) and the point-source object (PS). The right panel is the velocity dispersion map derived from the \ha\ emission line. The circles are the same as in the middle panel.}
\label{velocity_maps}
\end{figure*}

Figure~\ref{velocity_maps} shows the radial velocity map (left and middle panel) and the velocity dispersion map (right panel) of the \ha\ emission line. Single sincgauss fits were used to obtain these maps regardless of possible line-splitting. In the velocity field, the fit traces the velocity peak at each pixel. This gives a general sense of the kinematics of the nebula. In the case of the velocity dispersion, the width of a single sinc-Gaussian is reported. This is useful even if there are multiple components, as the width should be larger if there are multiple unresolved components. The velocity dispersion map was corrected by instrumental and thermal widths, $\sigma_{\text{inst}} = \SI{68.5}{\kilo\metre\per\second}$ and $\sigma_{\text{th}} = \SI{8.5}{\kilo\metre\per\second}$, respectively. The latter value was estimated by assuming an electron temperature of \SI{8680}{\kelvin} \citep{esteban2009} in the expression $\sigma_{\text{th}} = (k_B T_e/m_{\ch{H}})^{1/2}$. Using a ``standard'' electron temperature of \SI{e4}{\kelvin} increases the thermal width to \SI{9.1}{\kilo\metre\per\second} which is negligible due to the quadratic sum. 

A close look at Figure~\ref{velocity_maps} reveals that the velocity map is correlated somewhat with the \ha\ flux distribution, while the dispersion map correlates with the location of the known cavities in NGC~604. From west to east, there is an increase in the radial velocity of $\sim$\SI{30}{\kilo\metre\per\second}, which might suggest that the nebula is rotating. Given that NGC~604 is not likely spherical, it instead could be an expanding structure approaching on the west and receding on the east. The velocity and velocity dispersion maps agree well with those found by \cite{kam2015}, but the spatial resolution of \ac{sitelle} allows us to reveal more detail, particularly of the cavities. 

Inspecting the rightmost panel of Figure~\ref{velocity_maps}, we note that regions having the largest values of $\sigma$ are those for which there are likely multiple unresolved velocity components. This map is then a useful tool to search for expanding structures. We find that cavities A and D display broad multiple profiles (\qtyrange[range-units=single]{40}{50}{\kilo\metre\per\second}) while cavities B and C as well as most of the nebula display comparably narrower profiles (\qtyrange[range-units=single]{20}{30}{\kilo\metre\per\second}). The dispersions found in cavities A and D are comparable to the expansion velocities found by \cite{yang1996}, $v_{\text{exp}} \sim \SI{40}{\kilo\metre\per\second}$. The dispersions found elsewhere are comparable to the contribution from gravitational broadening that \cite{yang1996} found. 

In order to investigate a possible correlation between different line ratios and the velocity features seen here, we compare the velocity and dispersion maps in Figure~\ref{velocity_maps} with the excitation-sensitive emission line ratios in Figure~\ref{excitation}, particularly the $\log(\text{\sii/\ha})$ and $\log(\text{\oiii/\hb})$ maps. The $\log(\text{\oiii/\hb})$ ratio is sensitive to the radiation field and metallicity, while $\log(\text{\sii/\ha})$ is sensitive to shocks. Comparing these excitation maps to the velocity dispersion map, we see that the \ac{snr} M33-16 stands out in both the dispersion and $\log(\text{\sii/\ha})$ maps. 

Other high-dispersion features do not exhibit similarly high values of $\log(\text{\sii/\ha})$ that are indicative of shocks. For instance, cavities A and D show relatively low values of $\log(\text{\sii/\ha})$, but higher values of $\log(\text{\oiii/\hb})$. The low $\log(\text{\sii/\ha})$ values in this cavity do not suggest that it is powered by \acp{snr} (although \acp{snr} in superbubbles would be hard to detect until the shocks have hit the superbubble walls; \citealt{chu1990}). Instead, this area appears to be expanding; the velocity map shows a dipole of blueshifted and redshifted portions of the cavity. Most of the ionizing O-type and \ac{wr} stars are located below this cavity in cavity A, and it could be their stellar winds powering the dynamics in this cavity \citep{tullmann2008}.

Finally, there is a peculiar portion of gas to the southeast that exhibits coherent velocities. This region exhibits mild velocity dispersions but has $\log(\text{\sii/\ha})$ values consistent with the bulk of NGC~604, so shocks are not likely driving this motion. Additionally, the low $\log(\text{\oiii/\hb})$ values indicate that there is not a population of ionizing stars with strong stellar winds driving the motion either. \cite{maizapellaniz2004} suggested, based on BPT diagrams, that the nebula is density-bounded to the west. Perhaps this is a resulting champagne flow \citep{tenoriotagle1979}, where stellar feedback is blowing out material along the path of least resistance.

%%% ^ add citation to champagne flow

\section{The Importance of Spatially-Resolved Data}\label{sec:important}

% Does R23 work reasonably well if you only have partial sampling?
% Does anything not work at all?

The observations presented here of NGC~604 provide insight into this \ac{ghr} at \SI{3}{\parsec} scales, revealing a complex structure characterized by changes in ionization and affected by internal feedback. What do we gain from observing \acp{ghr} on such fine spatial scales? The first and most obvious gain is that we can observe the complex structure of the ionized gas in relation to its morphology and the physical feedback effects of the stars. For example, we have found two individual sources within the \hii\ region, a known \ac{snr} M33-16 and a mysterious point source. Both of these only contribute $\sim$\SI{1}{\percent} of the light to the integrated spectrum, and their presence would be missed entirely without spatially-resolved information. Using the emission lines, we have found arc structures indicative of shocks, changes in the ionization state of the gas, and potential density fluctuations. 

Despite the detection of shocked features, our emission line diagnostics show that the gas is predominantly photoionized. We would expect to see shock excitation given the young, intense star formation and consequently outflowing gas. However, when we analyze the shock-sensitive \sii/\ha\ maps, we do not find regions that appear particularly shock-excited, suggesting that stellar winds and \aclp{sne} have not yet come to dominate the kinetic energy of the \ac{ghr} \citep{tenoriotagle1996}. To detect the shocks, we likely require higher spectral resolution data. The SN3 data cube used a resolution of only $R = 2900$ corresponding to a $\Delta v \approx \SI{100}{\kilo\metre\per\second}$. Previous studies have measured line splitting occurring at $\Delta v \simeq \SI{60}{\kilo\metre\per\second}$ \citep{tenoriotagle2000} requiring $R \gtrsim 5000$. With such data, one could decompose the emission line components to obtain separate and more accurate line ratios. This technique is especially useful for isolating shock-excited emission \citep[e.g.,][]{rich2011,rich2014}, which, in the present study, would otherwise be lost among more dominant components from photoionization. 

\begin{figure}
\includegraphics[width=\columnwidth,keepaspectratio]{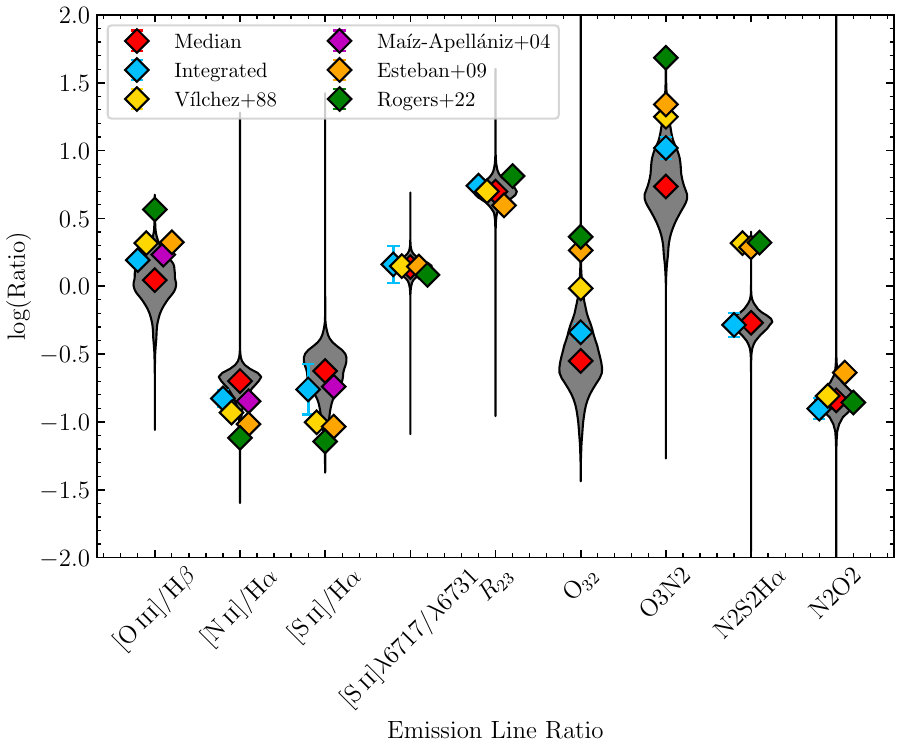}
\caption{Violin plots comparing the effects of resolution on particular line ratios. The gray violins show the distributions observed in the maps. The diamonds indicate integrated emission line ratios observed in this paper and from the literature \citep{vilchez1988,maizapellaniz2004,esteban2009,rogers2022} indicated in the legend; the blue diamond is the integrated value reported in Section~\ref{sec:integrated} while the red diamond is the median. The diamonds have been horizontally offset by arbitrary amounts in some cases to aid readability.}
\label{comparison}
\end{figure}

%%% Put these through calibrations and see how big the variations are

The detailed chemical and ionization structure observed here would naturally be lost if we imaged NGC~604 on larger spatial scales or with an integrated spectrum. We illustrate the latter in Figure~\ref{comparison}, which shows several of the emission line ratios discussed here. The violin plots display the observed distribution in each map, while the diamonds represent either the integrated spectrum presented here (blue diamond) or those reported in the literature. We see that the line ratios showing the strongest variations are those that are excitation- or ionization-sensitive, i.e., \oiii/\hb, \nii/\ha, \sii/\ha, O$_{32}$, and O3N2. The largest variations between the median and the integrated spectra are in O$_{32}$ and O3N2 of up to \SI{0.95}{\dex}. Meanwhile, the abundance-sensitive line ratios of $R_{23}$ and N2O2 fair much better, with a \SI{0.1}{\dex} and \SI{0.2}{\dex} variation, respectively.  As mentioned in Section~\ref{sub:strong_ratio_maps}, this manifests the robustness of these two indices against ionization variations. Interestingly, the density-sensitive \sii\ ratio shows the least variation between methods, indicating that integrated or long-slit spectra are sufficient to measure the electron density of an \hii\ region, even in \acp{ghr}. The cause of the consistent difference in the N2S2\ha\ index is unknown, which cautions against using this index. 

How do these line ratio variations translate into abundance and ionization parameter variations? To answer this question, we applied the abundance and ionization calibrations listed in Section~\ref{sub:oxy}, applied them to the spectra of \cite{vilchez1988}, \cite{esteban2009}, and \cite{rogers2022}, and compared them to our integrated abundances and ionization parameters reported in Table~\ref{int-ratios}. For all of the abundance indicators, we found only variations on the order of \SI{0.2}{\dex}, \added{comparable to the statistical uncertainties of the individual calibrators}. The only exception was for the $R_{23}$ calibration of \cite{pilyugin2005}, where the variations amounted to a maximal difference of \SI{0.35}{\dex}. The large difference found here is likely explained by the empirical nature of this calibration; strictly speaking, only data similar to the data used to construct the calibration should be applied. 

However, the largest variations were found in the ionization parameter. The largest difference from our integrated value was between \cite{esteban2009} and \cite{rogers2022} of a $\sim$\SI{0.8}{\dex} difference. Interestingly, the O3N2 \cite{marino2013} calibration does not result in a similarly large abundance variation despite its sensitivity to the ionization state. These three long-slit spectra likely target regions of NGC~604 with similar ionization states and mildly different abundances, resulting in a smaller range of abundances. Thus, while it seems that abundances can still be robustly estimated with spatially-limited data, measurements of the ionization parameter require spatially-resolved data to account for variations. 

Finally, we issue a word of caution towards other potential spatially-resolved observations, such as those collected with \acp{ifu} on \emph{JWST} or the SDSS Local Volume Mapper \citep{drory2024}. Strong-line abundance calibrations cannot be used on the individual spaxels of spatially-resolved data. Many of these calibrations are statistical relations between integrated \hii\ regions or whole galaxies \citep[e.g.,][]{pettini2004,pilyugin2005,marino2013}. Others are based on theoretical models that assume photoionization is the dominant excitation process \citep[e.g.,][]{mcgaugh1991,kobulnicky2004,dopita2016,kewley2019}; photoionization dominance may not be the case on small scales. Additionally, these calibrations work because they are underpinned by well-tested assumptions about how the oxygen abundance changes with the ionization parameter \citep[e.g.,][]{ji2022,garner2025} and the nitrogen abundance \citep[e.g.,][]{nicholls2017,berg2020}. Both assumptions only hold for integrated \hii\ regions, not for individual spaxels of a single \hii\ region such as the case here with NGC~604. Thus, while one can input spaxel line ratios into these calibrations and receive an oxygen abundance, the resulting oxygen abundance is meaningless and interpretations should be taken with extreme caution.

\section{Conclusions}\label{sec:conclusion}

In this work, we have analyzed the physical properties of the giant \hii\ region NGC~604, the most luminous and massive \ac{ghr} in the Local Group after 30~Doradus. We acquired the data with the \ac{cfht} imaging Fourier transform spectrometer \ac{sitelle} \citep{drissen2019}, which provides a spatial resolution of $\sim$\SI{3.2}{\parsec} and a spectral resolution ranging from $R = 1020$ in the blue to $R = 2900$ in the red. Since NGC~604 is located in the larger Field 1 of M33 \citep{tuquet2024,duartepuertas2024} with an area of \SI{121}{\square\arcmin}, we were able to study the whole spatial extent of NGC~604 down to a flux level of \SI{2e-17}{\erg\per\second\per\square\centi\metre} or a total area of \SI{1.825}{\square\arcmin}. This comprehensive coverage allowed us to fully explore the excitation, ionization, and potential abundance variations across the entirety of the nebula for the first time. Our main conclusions are summarized below. 

\begin{enumerate}

% Add a line about the correspondence between Balmer & PAH
\item Using the integrated spectrum of NGC~604, which shows strong optical lines as expected, we estimated the dust extinction ($A_V = 0.3$), the \ac{sfr} (\SI{0.018}{\solarmass\per\year}), the mass of the ionized gas (\SI{9e5}{\solarmass}), and the stellar mass (\SI{2.3e5}{\solarmass}), all comparable to previous imaging studies. We also utilized strong-line ratios to investigate the integrated excitation and abundance properties of NGC~604. These confirmed that NGC~604 is primarily excited by OB stars and has mildly subsolar oxygen abundances and mild ionization parameter. 

\item Turning to the spatially-resolved data, we constructed maps of several parameters and line ratios. Investigating the maps of excitation-sensitive line ratios (\nii/\ha, \sii/\ha, and \oiii/\hb) shows a bright, high-excitation core near the central stellar cluster surrounded by a dim, low-excitation halo. Several discrete sources, which are undetectable in the integrated spectrum, are easily found in these maps, such as \ac{snr} M33-16 \citep[e.g.,][]{dodorico1980} and a mysterious point source embedded in the nebula to the south. Constructing the standard BPT diagrams, we see that our integrated spectrum is more representative of the entire excitation structure than previous long-slit spectroscopic measurements. 

\item We then turn our attention to how line ratios usually used to measure abundance and ionization in integrated spectra of \hii\ regions vary across NGC~604, a first for this nebula. Using several strong-line ratios, including the $R_{23}$, O$_{32}$, O3N2, N2S2\ha, and N2O2, we see strong variations across the region. To interpret these variations, we construct ``residual'' maps by subtracting the integrated line ratio from each map. After accounting for data uncertainties, we find that most of the variations are \added{statistically insignificant}, and those that remain likely result from changes in ionization properties rather than chemical inhomogeneities. We warn the reader about potential secondary dependencies for each strong-line ratio that might bias straightforward interpretation. 

\item Finally, we present a ``holistic'' view of the ionized gas kinematics of NGC~604 across the entire nebula. We largely confirm previous studies, particularly those that investigated the kinematics of the four large cavities. For instance, cavities A and D have large velocity dispersions, likely indicative of split-line profiles that are unresolved in our data. We connect the dispersions to our excitation-sensitive line ratio maps, indicating that most of the dispersion does not appear to be driven by shocks but rather by stellar winds and photoionization. 

\end{enumerate}

Overall, our results are consistent with previous studies of NGC~604; however, our study expands the view and provides new insights into its abundance and ionization structure. We stress the importance of spatially-resolved studies, especially of nearby \hii\ regions, in order to explore the full ionization and abundance structure, which will aid our interpretation and assumptions about extragalactic \hii\ regions where we do not necessarily have spatially-resolved information. As a concluding remark, this research is only a first step. \ac{signals} has many more giant \hii\ regions like NGC~5461 and NGC~5471 in M101, which are the two brightest after NGC~604, and many others in Local Group galaxies such as IC~1613 and NGC~4395. However, these are at lower spatial scales due to their larger distances but have been observed at higher spectral resolutions. A future paper will compile a similar spatial analysis of these giant \hii\ regions, expanding our insights into the ionization and abundance structure of these complex regions.

\begin{acknowledgments}
R.G.\ would like to thank Thomas Martin for his help in fully utilizing \textsc{orcs} to fit the spectra presented in this paper. The authors would like to thank the anonymous referee for their comments which greatly helped the consistency and readability of the paper. This paper is based on data obtained for \ac{signals}, a large program conducted at the \acf{cfht}, which is operated by the National Research Council of Canada, the Institut National des Sciences de l'Univers of the Centre National de la Recherche Scientifique of France, and the University of Hawaii. The observations were collected with \ac{sitelle}, a joint project of Universit\'{e} Laval, ABB, Universit\'{e}  de Montr\'{e}al and the \ac{cfht}, with support from the Canada Foundation for Innovation, the National Sciences and Engineering Research Council of Canada (NSERC), and the Fonds de Recherche du Qu\'{e}bec--Nature et Technologies (FRQNT). The authors wish to recognize and acknowledge the very significant cultural role that the summit of Mauna Kea has always had within the indigenous Hawaiian community. We are most grateful to have the opportunity to conduct observations from this mountain. C.R.\ and L.D.\ acknowledge financial support from the National Science and Engineering Research Council of Canada and the Fonds de Recherche du Qu\'{e}bec (\url{https://doi.org/10.69777/283645}). C.M.\ acknowledges the support of grant UNAM/DGAPA/PAPIIT IG101223.

This work has made use of data from the European Space Agency (ESA) mission \emph{Gaia}, processed by the \emph{Gaia} Data Processing and Analysis Consortium (DPAC). Funding for the DPAC has been provided by national institutions, in particular the institutions participating in the \emph{Gaia} Multilateral Agreement. This research has made use of the Astrophysics Data System, funded by NASA under Cooperative Agreement 80NSSC21M00561. Some of the data presented in this paper were obtained from the Mikulski Archive for Space Telescopes (MAST). STScI is operated by the Association of Universities for Research in Astronomy, Inc., under NASA contract NAS5-26555. Support for MAST for non-HST data is provided by the NASA Office of Space Science via grant NNX13AC07G and by other grants and contracts. The \emph{JWST} data used here may be obtained from the MAST archive at \dataset[doi:10.17909/chgn-pf19]{https://doi.org/10.17909/chgn-pf19}.
\end{acknowledgments}

\facility{CFHT}

\software{\texttt{Astropy} v7.0.0 \citep{astropy1,astropy2,astropy3}, \texttt{Matplotlib} v3.9.2 \citep{hunter2007}, \texttt{NumPy} v1.26.4 \citep{harris2020}, \texttt{SciPy} v1.13.1 \citep{virtanen2020}, \textsc{orb} v3.0 \citep{martin2015}, \textsc{orcs} v3.0 \citep{martin2015,martin2020}, \texttt{reproject} v0.14.1 \citep{robitaille2020}, \texttt{PyNeb} v1.1.23 \citep{luridiana2015}, \texttt{regions} v0.10\footnote{\url{https://reproject.readthedocs.io/en/stable/}}, \texttt{multicolorfits} v2.1.2 \citep{cigan2019}, \texttt{cmcrameri} v1.8 \citep{crameri2018}}

\bibliography{n604.bib}
\bibliographystyle{aasjournal.bst}

\appendix
\section{Flux Maps}
\begin{figure*}
\centering
\includegraphics[height=0.8\textwidth,keepaspectratio,angle=270]{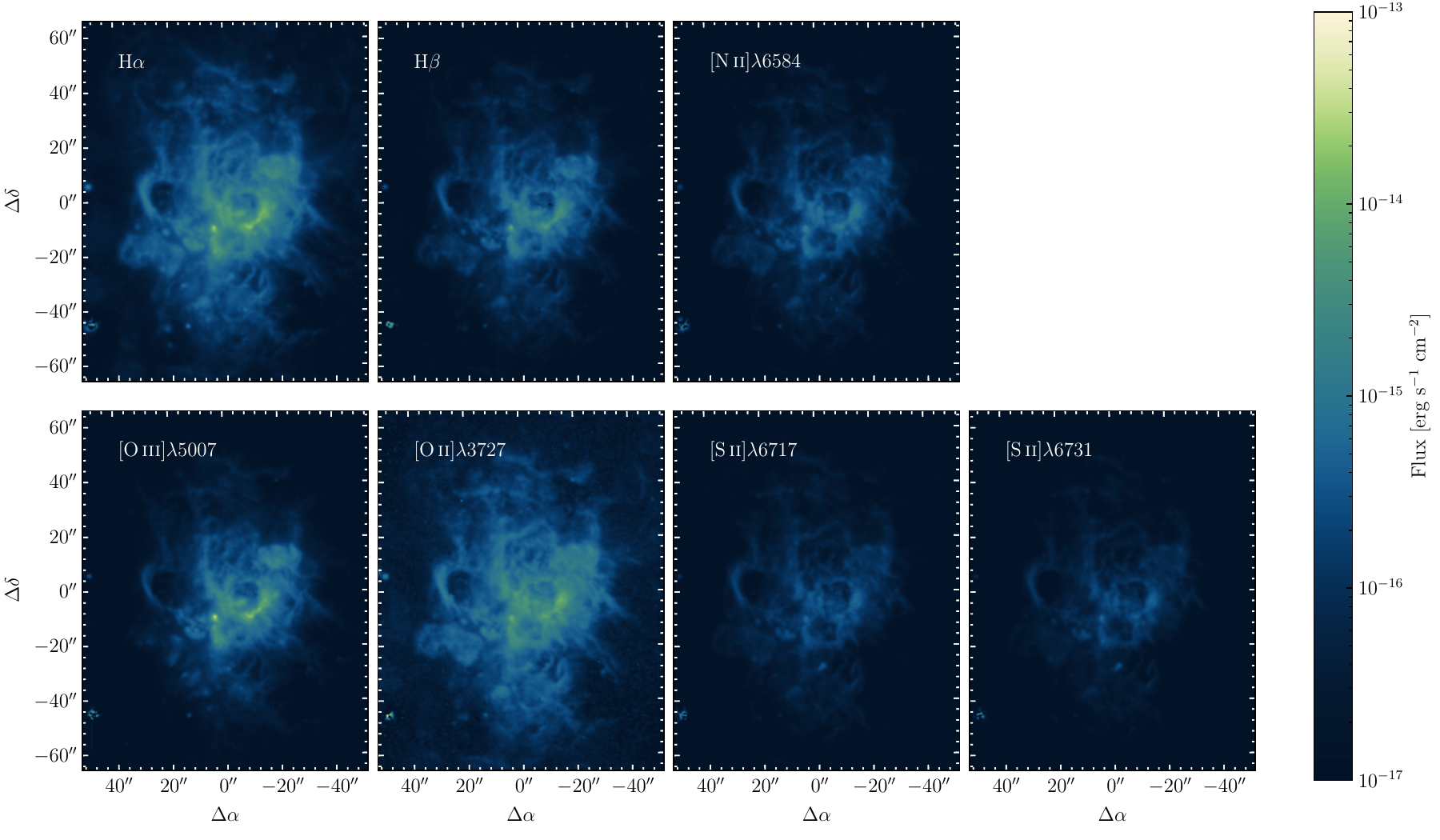}
\caption{Flux maps of the emission lines at NGC~604. Each line is labeled at the top left and are presented using the same scaling. An internal extinction correction has not been applied. North is up and east is to the left.}
\label{emission_lines}
\end{figure*}

The flux maps for all of the emission lines used in this paper are shown in Figure~\ref{emission_lines}. Emission lines from the same filter have been fit simultaneously using routines in \textsc{orcs} applied to all the spaxels individually. The subtraction of the galaxy stellar population spectrum along the line-of-sight for each spaxel was done (see, e.g., \citealt{tuquet2024,duartepuertas2024} for more information). The correction for internal extinction has not been applied to this figure. All spaxels have been kept regardless of signal-to-noise. 

\end{document}